\documentclass[a4paper,onecolumn,10pt,accepted=2019-05-03, issue=1, volume=1, shorttitle=papers/compositionality-1-1]{compositionalityarticle}
\pdfoutput=1
\usepackage[utf8]{inputenc}
\usepackage[english]{babel}
\usepackage[T1]{fontenc}
\usepackage{amsmath}
\usepackage{hyperref}
\usepackage[numbers,sort&compress]{natbib}

\usepackage{amsfonts}
\usepackage{enumerate}
\usepackage{physics}
\usepackage{mathtools}
\usepackage{xspace}
\usepackage[toc,page]{appendix}
\usepackage{amsthm}

\usepackage{tikz}
\usepackage{tikz-cd}
\usetikzlibrary{arrows, matrix}

\usepackage{enumitem}

\DeclarePairedDelimiter{\ceil}{\lceil}{\rceil}
\DeclarePairedDelimiter{\floor}{\lfloor}{\rfloor}
\DeclarePairedDelimiter{\inn}{\langle}{\rangle}

\newcounter{counter}
\numberwithin{counter}{section}
\newtheorem{theorem}[counter]{Theorem}
\newtheorem{proposition}[counter]{Proposition}
\newtheorem{definition}[counter]{Definition}
\newtheorem{lemma}[counter]{Lemma}
\newtheorem{remark}[counter]{Remark}
\newtheorem{example}[counter]{Example}
\newtheorem{corollary}[counter]{Corollary}
\newtheorem*{theorem*}{Theorem}
\newtheorem*{lemma*}{Lemma}
\newtheorem*{definition*}{Definition}

\newcommand{\rnk}{\text{rnk}\xspace}
\newcommand{\R}{\mathbb{R}}
\newcommand{\C}{\mathbb{C}}
\newcommand{\N}{\mathbb{N}}

\newcommand{\OUS}{\text{\textbf{OUS}$^\text{op}$}\xspace}
\newcommand{\EJA}{\text{\textbf{EJA}}\xspace}

\newcommand{\asrt}{\text{asrt}}
\newcommand{\pred}{\text{Eff}}
\newcommand{\id}{\text{id}}
\newcommand{\opp}{\text{op}}
\newcommand{\cl}[1]{\overline{#1}}
\newcommand{\im}[1]{\text{im}(#1)}

\usepackage[xcolor]{changebar}

\begin{document}

\title{An effect-theoretic reconstruction of quantum theory}
\date{}
\author{John van de Wetering}
\email{john@vdwetering.name}
\affiliation{Radboud University Nijmegen, Houtlaan 4, 6525 XZ Nijmegen, Netherlands}

\maketitle

\begin{abstract}
An often used model for quantum theory is to associate to every physical system a C$^*$-algebra. From a physical point of view it is unclear why operator algebras would form a good description of nature. In this paper, we find a set of physically meaningful assumptions such that any physical theory satisfying these assumptions must embed into the category of finite-dimensional C$^*$-algebras. These assumptions were originally introduced in the setting of \emph{effectus theory}, a categorical logical framework generalizing classical and quantum logic. As these assumptions have a physical interpretation, this motivates the usage of operator algebras as a model for quantum theory.

In contrast to other reconstructions of quantum theory, we do not start with the framework of generalized probabilistic theories and instead use \emph{effect theories} where no convex structure and no tensor product needs to be present. The lack of this structure in effectus theory has led to a different notion of \emph{pure} maps. A map in an effectus is pure when it is a composition of a \emph{compression} and a \emph{filter}. These maps satisfy particular universal properties and respectively correspond to `forgetting' and `measuring' the validity of an effect.

We define a \emph{pure effect theory} (PET) to be an effect theory where the pure maps form a dagger-category and filters and compressions are adjoint. We show that any convex finite-dimensional PET must embed into the category of \emph{Euclidean Jordan algebras}. Moreover, if the PET also has monoidal structure, then we show that it must embed into either the category of real or complex C$^*$-algebras, which completes our reconstruction. 
\end{abstract}

\section{Introduction}
Quantum theory may be taken to be the part of quantum physics that deals with physical systems in an abstract way, without referencing any concrete physical implementation. A \emph{qubit} for instance, is not any particular physical system, but rather is a mathematical description that applies to a class of existing systems.
A common way to model a system in quantum theory is to associate a (finite-dimensional) \emph{C$^*$-algebra} to it, such as representing a qubit by $M_2(\C)$, the set of complex $2\times 2$ matrices. It is not clear at all at first glance why we should model physical systems using C$^*$-algebras. Could it not be that some other model would work better? Or can we maybe find a reason why we need to use C$^*$-algebras?

Generalised Probabilistic Theories (GPTs) were introduced as a general framework that should be able to fit any physical theory. They are a widely used framework for studying what properties are specific to quantum theory and which are generic~\cite{barrett2007information,barrett2017localstoch,chiribella2010probabilistic,barnum2007generalized,short2010entropy}. One particular direction of research in which the usage of GPTs is widespread is in studying physical principles from which the structure of quantum theory, C$^*$-algebras, can be derived; a process known as \emph{reconstructing quantum theory}~\cite{hardy2001quantum,masanes2011derivation,masanes2014entanglement,chiribella2011informational,dakic2009quantum,hardy2011reformulating,wilce2018royal,wetering2018sequential,gunson1967algebraic}\footnote{Although some reconstructions use a (slightly) different framework~\cite{barnum2014higher,selby2018reconstructing,tull2016reconstruction,fivel2012derivation,hohn2017toolbox,goyal2010information}.}. Such physical principles tell us then that any physical theory that is `well-behaved' enough to have those principles, must be describable by C$^*$-algebras or related structures.
Using GPTs, several `foil' theories have also been found that can be contrasted with quantum theory~\cite{spekkens2007toy,popescu2014nonlocality,chiribella2016purity,gogioso2017fantastic}, which in combination with the physical principles found in reconstructions of quantum theory allow us to see more clearly how quantum theory is special and how it is not.

A different framework for studying quantum theory from a broad point of view is that of \emph{effectus theory}~\cite{cho2015introduction}. This is a categorical framework for studying logic in both classical and quantum systems. While notions from GPTs like \emph{states} and \emph{effects} are still applicable in an effectus, it generalises the notion of probability by replacing the familiar convex structure by \emph{effect algebras}. As a result, a wider array of theories where scalars have a more complicated structure are included in its definition, which in principle could include examples like the space-like quantum theory in Ref.~\cite{moliner2017space} where the scalars are the unit interval of a commutative C$^*$-algebra or the quantum theory based on non-standard analysis of Ref.~\cite{gogiosotowards} where the scalars are the hyperreal numbers.

As opposed to GPTs, effectus theory started from the ground up in category theory, and in the most general setting does not have a notion of tensor product. As a result, new concepts were developed that could be used in this general setting. In particular, \emph{quotients} and \emph{comprehensions}~\cite{cho2015quotient} were found to provide a rich structure that allows one to give a new definition of purity for maps~\cite{bramthesis,basthesis}. This definition has proved to be quite fruitful in studying purity in settings beyond finite-dimensional quantum theory~\cite{westerbaan2016paschke,westerbaan2016universal,westerbaan2018puremaps}, as their compositional structure seems to be more well-behaved than other ones in use (see Section~\ref{sec:purity} for an extended discussion on different notions of purity found in the literature). 
Additionally, by using the notion of an \emph{image} of a map, a new definition of \emph{sharpness} can be given for effects.

As the definition of an effectus is quite involved, we will instead use the weaker notion of an \emph{effect theory}, to be defined later. The main object of interest in this paper is the following:

\begin{definition*}
    A (monoidal) \textbf{pure effect theory} (PET) is a (monoidal) \textbf{effect theory} where
    \begin{itemize}
        \item All effects have filters and compressions.
        \item All maps have images.
        \item The negation of a sharp effect is sharp.
        \item The pure maps form a (monoidal) dagger-category.
        \item Filters and compressions of a sharp effect are adjoint.
        \item Compressions of sharp effects are isometries.
    \end{itemize}
\end{definition*}

The definitions of all the terms can be found in Section~\ref{sec:purity}. Therein it is also illustrated how these, quite mathematically stated, `axioms' make intuitive sense to hold for a theory of physics.

The main example of a pure effect theory is the category of \emph{Euclidean Jordan algebras} (EJAs) with positive sub-unital maps~\cite{westerbaan2018puremaps}. EJAs were one of the first generalisations of quantum theory to be studied~\cite{jordan1933}. Examples of EJAs include matrix algebras of real, complex or quaternionic self-adjoint matrices (and in fact, this list almost completely exhausts the list of examples~\cite{jordan1993algebraic}). The example of EJAs is in a sense the most general one, which brings us to the main result of this paper:
\begin{theorem*}[informal]
    A (non-monoidal) GPT that is a pure effect theory embeds into the category of Euclidean Jordan algebras with positive sub-unital maps.
\end{theorem*}

\noindent It should be noted that finding suitable tensor products for EJAs is still an open problem~\cite{barnum2016composites}, which is why we do not require a tensor product in the above theorem. In fact, it seems that the compositional structure of systems sets regular quantum theory apart from the more general EJAs:

\begin{theorem*}[informal]
    A GPT that is a monoidal pure effect theory embeds into either the category of complex C$^*$-algebras, or the category of real C$^*$-algebras with completely positive sub-unital maps.
\end{theorem*}
\noindent The reason we get both real and complex quantum theory in the end, is because we do not require \emph{local tomography}~\cite{hardy2012limited} for monoidal pure effect theories, which is a common way to distinguish the two.

As a monoidal PET in the standard probabilistic setting corresponds to quantum theory in this manner, we can interpret a PET where the probabilities take values in a different set to be a foil for quantum theory in a universe where the rules of probability work differently. This allows for a different perspective than the foil theories developed in the context of GPTs where the convex probabilistic structure is always present.

The paper is structured as follows. In Section~\ref{sec:framework} we introduce our framework of \emph{effect theories}, which is a stripped-down version of an effectus. We also introduce \emph{operational} effect theories, which are our version of GPTs. 
Section~\ref{sec:purity} contains an extended discussion on the different notions of purity that can be found in the literature, and why they are not suitable for our use case, before introducing the version of purity used in effectus theory and the related notions of filters and compressions.
In Section~\ref{sec:firsttwo} we start proving general properties of PETs before transitioning to operational PETs in Section~\ref{sec:opeffect} where their relation to Jordan algebras is established. Finally, in Section~\ref{sec:quantum} we study monoidal operational PETs.

\subsection{Comparison to other reconstructions}
There is a variety of reconstructions of quantum theory using a variety of axioms. Nevertheless, some common themes amongst these axioms can be found. 

First off is some kind of \emph{preservation of purity} demand~\cite{chiribella2015operational,barnum2014higher,hardy2011reformulating,gunson1967algebraic,tull2016reconstruction,selby2018reconstructing} which states that a composition of pure maps (where the definition of pure varies) should again be a pure map. In our setting this corresponds to the requirement that the pure maps should form a category.

Next up is the existence of certain types of filters or the ability to restrict systems to subsystems based on the support of effects or states~\cite{chiribella2011informational,hardy2011reformulating,wilce2018royal,tull2016reconstruction,barnum2014higher,gunson1967algebraic}, which in this reconstruction takes the form of asking filters and compressions to exist.

An almost universally used type of axiom that is not present in this paper, is the requirement of some kind of \emph{symmetry} or \emph{convertability} between pure states, i.e.\ the requirement that a certain amount of reversible transformations must exist~\cite{barnum2014higher,hardy2001quantum,hardy2011reformulating,dakic2009quantum,masanes2014entanglement,krumm2017thermodynamics,chiribella2011informational,tull2016reconstruction,selby2018reconstructing,hohn2017toolbox}. \footnote{In fact, the only reconstructions except for the authors previous work~\cite{wetering2018sequential} that the author is aware of that do not say anything explicitly about reversible transformations are Refs.~\cite{gunson1967algebraic,wilce2018royal} which both require the strong assumption of self-duality of the space from the start.}

Our axiom stating that compressions of sharp effects are isometries is closely related to those compressions being \emph{dagger kernels}~\cite{heunen2010quantum}. In Ref.~\cite{tull2016reconstruction} the existence of dagger kernels is assumed as an axiom. The fact that a compression has a one-sided inverse is also related to the \emph{ideal compression} axiom of Ref.~\cite{chiribella2011informational}. See \cite[Theorem 4.21]{tull2019phdthesis} for a more formal version of this relation.

The dagger structure axioms for pure maps that we require have no direct comparison in the literature, but it should be noted that in Refs.~\cite{tull2016reconstruction,selby2018reconstructing} they require that \emph{all} maps have a dagger, and particularly the \emph{sharp dagger} of Ref.~\cite{selby2018reconstructing} is used in a similar way as in this paper, namely to derive \emph{symmetry of transition probabilities} (Proposition~\ref{prop:symoftrans}). The existence of images has no counterpart in the literature\footnote{In Ref.~\cite{tull2016reconstruction} maps called `images' are used, but they are not related to the images of this paper in any direct manner.}.

It is well-established that the existence of a tensor product satisfying \emph{local tomography} allows one to distinguish arbitrary Euclidean Jordan algebras from complex C$^*$-algebras~\cite{tull2016reconstruction,selby2018reconstructing,hohn2017toolbox,wilce2018royal,wetering2018sequential} (although some other papers use local tomography in a more substantial way as part of their derivation~\cite{hardy2001quantum,chiribella2011informational}). In this paper we use the assumption of the existence of a well-behaved tensor-product in order to force our systems to be real or complex C$^*$-algebras, a dichotomy that also been found before~\cite{hanche1985jb,barnum2016composites}.

Finally, while this paper uses ideas of a general categorical nature, core methods of the proof only work in the setting of GPTs. This is in contrast to Ref.~\cite{tull2016reconstruction} where the structure of the real numbers is only needed at the very end, and the majority of the work is done in the abstract setting of category theory.

\section{Effect Theories}\label{sec:framework}
We start by describing the framework we adopt. This framework is inspired by generalized probabilistic theories, but is more general because we don't yet assume a convex structure on our transformations.

The idea of the framework is to model in an abstract way the things you could possibly do in a lab. To this end our most basic notion is that of a \emph{system}. This represents any kind of physical object that you could do experiments on. We will denote systems by capital letters $A,B,C,\ldots$.

We might be able to influence or change a system $A$ in some manner to get a different system $B$. We will denote such transformations by $f:A\rightarrow B$.

Of course we can do transformations one after another, which we will denote by $g\circ f$, and this should satisfy the obvious associativity condition $h\circ(g\circ f) = (h\circ g)\circ f$. Combined with the fact that we of course have the `trivial' transformation $\id: A\rightarrow A$ which simply means we do nothing, we have found the familiar notion of a \emph{category}.

Next we will assume that we have some kind of special system $I$ that denotes the `empty' or `trivial' system. A transformation $\omega:I\rightarrow A$ is then a procedure that creates a system from nothing, i.e.\ it is a preparation. We will call such transformations \emph{states}. They represent the different ways in which a system $A$ can be prepared. We define the \emph{state space} of $A$ to be St$(A):=\{\omega: I\rightarrow A\}$.

Dually, the transformations $q:A\rightarrow I$ are the ways in which the system $A$ can be destroyed. We call these transformations \emph{effects} and they model the different types of measurements you can perform on a system $A$. We define the \emph{effect space} of $A$ to be Eff$(A):=\{q:A\rightarrow I\}$.

The composition of a state with an effect results in a transformation $q\circ\omega: I\rightarrow I$. Such a transformation from the unit system to itself is called a \emph{scalar}. 

A generalized probabilistic theory fits in this framework. In that case, the states and effects have a convex structure, and the scalars are real numbers from the unit interval. The scalar $q\circ \omega$ is then the probability that a measurement $q$ returns true on a state $\omega$. 
Instead of this identification of the scalars with the real numbers and a convex structure, we will require a weaker sort of structure, known as an \emph{effect algebra}. The effect algebra structure will allow us to consider sums and negations of effects.

\begin{definition}
	An \textbf{effect algebra} $(E,+,1,0,\perp)$ is a set $E$ equipped with a partially defined commutative, associative addition operation $+$ (so if $x+y$ is defined then $y+x$ is also defined and $x+y=y+x$. Similar rules hold for associativity), such that 
    \begin{itemize}
        \item For all $x$, $x+0=x$,
        \item When $x+1$ is defined, then $x=0$,
        \item For all $x$ there exists a unique $y$ such that $x+y=1$. We denote this $y$ by $x^\perp$ or $1-x$ and it is called the \textbf{complement} of $x$.
    \end{itemize}
\end{definition}
\begin{remark}
    In an effect algebra the addition is automatically cancellative: $x+y=x+z\implies y=z$. As a result, it has a natural partial order defined as $x\leq y\iff \exists z: x+z=y$. This $z$ is unique and will be denoted by $y-x$, so that $x+(y-x) = y$. The complement reverses the order: $x\leq y \iff y^\perp \leq x^\perp$.
\end{remark}

Effect algebras were originally studied in for instance \cite{foulis1994effect,bennett1997interval} as generalisations of the space of effects in a quantum system $M_n(\C)$, those being Eff$(M_n(\C)):= \{ E\in M_n(\C)~;~ 0\leq E\leq 1\}$. This set forms an effect algebra where addition of effects $E$ and $F$ is defined as $E+F$ whenever $E+F\leq 1$. The complement is defined as $E^\perp = 1-E$. The induced order is the standard one for (positive) matrices. There are many more examples of effect algebras. In particular, the unit interval of any ordered vector space is an effect algebra. Viewing effects as possible measurements, the complement $E^\perp$ of an effect $E$ can be seen as negating the measurement outcomes of $E$, and then of course $E+E^\perp = 1$ gives the measurement that always returns true.

We now have the necessary ingredients to give the formal definition of our framework.

\begin{definition}
	An \textbf{effect theory} is a category $C$ with a designated object $I$ that we call the \textbf{trivial system} such that for all objects $A$, its \textbf{effect space} Eff$(A):=\{f:A\rightarrow I\}$ is an effect algebra, and such that every map $g:B\rightarrow A$ preserves addition: $0\circ g = 0$ and $(p+q)\circ g = (p\circ g)+(q\circ g)$ whenever $p+q: A\rightarrow I$ is defined as an effect.
\end{definition}

This definition is very general and includes a multitude of structures. In particular, any effectus in partial form~\cite{cho2015introduction} is an effect theory. Examples of effectuses include the category of sets with partial maps, the category of von Neumann algebras with normal positive contractive maps and the category of order unit groups. In fact, any biproduct category where a notion of `contractivity' for a map can be defined results in an effectus~\cite{cho2015introduction}. 
As the effects of an object form an effect algebra, we have for each object a special effect $1$. In the example of von Neumann algebras, this $1$ corresponds to the identity of the algebra. In causal generalized probabilistic theories, it corresponds to the unique \emph{causal} effect, also known as the \emph{(partial) trace}.


We will also need to be able to describe composite systems, i.e.\ given a couple of independent systems $A$ and $B$, it should be possible to describe them as forming one bigger system $A\otimes B$. The structure we wish to use for this is that of a \emph{monoidal category}.


\begin{definition}
    We call an effect theory $(C,I)$ \textbf{monoidal} when $C$ is a monoidal category such that the trivial system $I$ is also the monoidal unit and such that the tensor product preserves the effect algebra structure: $0\otimes q = 0$, $1\otimes 1 = 1$ and for all $q$, whenever $p_1+p_2$ is defined, $(p_1\otimes q)+(p_2\otimes q)$ is also defined and equal to $(p_1+p_2)\otimes q$.
\end{definition}

\subsection{Operational Effect Theories}

An important example of an effect theory concerns the category of \emph{order unit spaces}.

\begin{definition} \label{def:OUS}
    An \textbf{order unit space} (OUS) $(V,\leq,1)$ is a real vector space $V$ partially-ordered by $\leq$ with a special designated \textbf{order unit} $1$ such that for all $a,b,c\in V$ and $\lambda\in\R_{> 0}$:
    \begin{itemize}
        \item $a\leq b \iff a+c\leq b+c$,
        \item $a\leq b \iff \lambda a \leq \lambda b$,
        \item there exists $n\in \N$ such that $-n 1 \leq a \leq n 1$,
        \item if $a\leq \frac1n 1$ for all $n\in \N$ then $a\leq 0$
    \end{itemize}
\end{definition}

\begin{definition}
    We call a linear map $f:V\rightarrow W$ between order unit spaces \textbf{positive} when $v\geq 0\implies f(v)\geq 0$. It's \textbf{sub-unital} when $f(1)\leq 1$. We denote the category of order unit spaces with positive, sub-unital maps by \textbf{OUS}.
\end{definition}

The category \textbf{OUS}$^\opp$ is an effect theory. The states of this category correspond to positive sub-unital maps $\omega: V\rightarrow \R$, while the effects are positive sub-unital maps $\hat{q}:\R \rightarrow V$. These correspond in the obvious manner to elements $q\in [0,1]_V = \{v\in V~;~0\leq v \leq 1\}$.

The category \textbf{OUS}$^\opp$ is important, because it is in a sense the most general example of what we call an \emph{operational effect theory}. Such categories roughly correspond to the widely studied \emph{generalized probabilistic theories}. Before we give the definition of an operational effect theory, we must introduce a few other concepts.
\begin{definition}
    In an effect theory,
    \begin{itemize}
        \item We say the unital states \textbf{order-separate} the effects when for every system $A$ and effects $p,q\in \pred(A)$ we have $p\leq q$ whenever $ p\circ \omega \leq q\circ \omega $ for all $\omega\in $ St$(A)$ with $1\circ\omega = 1$.
        \item We say it satisfies \textbf{local tomography} when the effects separate the transformations: for all $f,g: B\rightarrow A$ we have $f=g$ whenever $p\circ f = p\circ g$ for all $p\in \pred(A)$.
        \item If it is monoidal we say it satisfies \textbf{tomography} when the effects monoidally separate the transformations: for all $f,g: B\rightarrow A$ we have $f=g$ whenever $p\circ (f\otimes \id_C) = p\circ (g\otimes \id_C)$ for all systems $C$ and $p\in \pred(A\otimes C)$.
    \end{itemize}
\end{definition}

\noindent Local tomography tells us that transformations are completely determined by what they do on effects. When monoidal structure is present in the theory, there are more ways to let a transformation interact with an effect, and hence we weaken the definition to (non-local) tomography.
The states order-separating the effects essentially tells us that there can be no `infinitesimal' effects (see Appendix~\ref{sec:opefftheory}). These properties allow us to relate effect theories to order unit spaces:

\begin{proposition}
    Let $\mathbf{E}$ be an effect theory where the scalars are the real unit interval, and where the unital states separate the effects. Then for all systems $A$ of $\mathbf{E}$, there exists an order unit space $V_A$, such that $\pred(A) \cong [0,1]_{V_A}$.
\end{proposition}
\begin{proof}
    This proof mostly follows Ref.~\cite{jacobs2016expectation}. For the details we refer to Appendix~\ref{sec:opefftheory}.
\end{proof}

The order unit space $V_A$ associated to $A$ has its effects associated one-to-one with those of $A$, but this is not necessarily true for the states. Any state $\omega:I\rightarrow A$ can be mapped to a state $\omega^*:V_A\rightarrow \R$ (see Theorem~\ref{theor:opefftheor} below), but not every state on $V_A$ necessarily comes from some state in the effect theory.

\begin{definition}
    Let $A$ be a system in an effect theory with real scalars where the unital states separate the effects and let $V_A$ be its associated effect space: $[0,1]_{V_A} \cong \pred(A)$.
    \begin{itemize}
    \item We say $A$ is \textbf{finite-dimensional} when $V_A$ is.
    \item We call $A$ \textbf{state-closed} when the collection of unital states St$_1(A)$ is closed as a subset of $V_A^*$ (with respect to the topology induced by the norm). 
    \item We call $A$ \textbf{scalar-like} when $\pred(A) \cong [0,1]$.
    \end{itemize}
\end{definition}

\begin{definition}
    A (monoidal) \textbf{operational effect theory} (OET) is a (monoidal) effect theory satisfying the following additional properties.
    \begin{enumerate}
        \item The set of scalars is the real unit interval: Eff$(I)=[0,1]$.
        \item The unital states order-separate the effects.
        \item All systems are finite-dimensional.
        \item All systems are state-closed.
        \item Every scalar-like system is isomorphic to the trivial system.
    \end{enumerate}
\end{definition}
Note that we do not require an OET to satisfy (local) tomography. As a result, this last condition is needed to prevent the case where we have multiple copies of the trivial system that are taken to be non-isomorphic, and can be seen as a very restricted form of tomography.

Let us compare this definition of an OET to that of a generalized probabilistic theory.
In both cases the scalars are of course the real numbers. 
The order-separation of the effects is similar to \emph{operational equivalence} in the literature on GPTs~\cite{chiribella2011informational}, but somewhat stronger, as they prevent the presence of infinitesimal effects. The effects forming effect algebras ensures that it is a \emph{causal} GPT, and furthermore, the effect algebra structure ensures that every effect $q$ has a complement $q^\perp$. Though these properties are not implied by the definition of a GPT, they are often assumed as a background assumption~\cite[Chapter 2]{tull2019phdthesis}. In the same way, finite-dimensionality and the closure of the state space are standard background assumptions~\cite{chiribella2011informational,barnum2014higher}.

If an OET satisfies local tomography, then all the information in the theory is captured by the structure of the effects:

\begin{theorem}\label{theor:opefftheor}
    Let $\mathbb{E}$ be an operational effect theory. Then there is a functor $F: \mathbb{E}\rightarrow \OUS$ such that $\pred(A)\cong [0,1]_{F(A)}$. Furthermore, this functor is faithful if and only if $\mathbb{E}$ satisfies local tomography.
\end{theorem}
\begin{proof}
    This proof mostly follows Ref.~\cite{jacobs2016expectation}. For the details we refer to Appendix~\ref{sec:opefftheory}.
\end{proof}

\subsection{Euclidean Jordan algebras}
\emph{Euclidean Jordan algebras} (EJAs) are a type of algebra originally intended to generalise the space of observables of a quantum system. 

\begin{definition}
    We call $(V,*,1)$ a \textbf{Jordan algebra} when $V$ is a real vector space with a bilinear commutative \textbf{Jordan product} $*:V\times V\rightarrow V$ that satisfies the \textbf{Jordan identity}: for all $a,b\in V$ $(a*a)*(b*a) = ((a*a)*b)*a$. A \textbf{Euclidean Jordan algebra} (EJA) is a finite-dimensional Jordan algebra equipped furthermore with an inner product satisfying for all $a,b,c\in V$ $\inn{a*b,c} = \inn{b,a*c}$ 
\end{definition}

Instead of introducing the full theory of EJAs, which is quite involved, we will use the Jordan-von Neumann-Wigner classification theorem to motivate why the EJAs are an interesting class of spaces, and to obviate the need for any in-depth development of the theory.

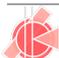
\begin{theorem}[Jordan-von Neumann-Wigner~\cite{jordan1993algebraic}]
    Any EJA can be written as a direct sum of the following \textbf{simple} EJAs:
    \begin{itemize}
        \item $M_n(F)^{sa}$, the set of $n\times n$ self-adjoint matrices over the field $F$ where $F$ is either the real numbers, the complex numbers or the quaternions. The Jordan product is given by $A*B := \frac12 (AB+BA)$ and the inner product is $\inn{A,B}:= \tr(AB)$.
        \item $H\oplus \R$ where $(H,\inn{\cdot,\cdot})$ is a real finite-dimensional Hilbert space. The product is defined by $(v,t)*(w,s) = (sv+tw,ts+\inn{v,w})$ and the inner product is $\inn{(v,t),(w,s)} = \inn{v,w}+ts$. These spaces are called \textbf{spin-factors}.
        \item The \textbf{exceptional} Jordan algebra $M_3(\mathbb{O})^{sa}$ of $3\times 3$ self-adjoint matrices over the octonions, with the Jordan product and inner product analogous to the first case.
    \end{itemize}
\end{theorem}
Of these possibilities, the spin-factors might seem strangest. Some of them should however be quite familiar. Any ordered vector space for which the state-space is affinely isomorphic to an $n$-dimensional ball is isomorphic to a spin-factor. In particular, the qubit system $M_2(\C)^{sa}$ is a spin-factor with a 3-dimensional Hilbert space given by the linear span of the Pauli matrices. The spin-factors can therefore be seen as generalisations of the qubit.

From this classification result, it should hopefully be clear why EJAs can be considered `quantum-like'. Beyond the usual complex C$^*$-algebraic framework, it also allows real and quaternionic systems. EJAs can be organised into a category.

\begin{definition}
    Let $V$ be an EJA. We write $a\geq 0$ for $a\in V$ when $\exists b\in V$ such that $a = b*b$. We call a linear map $f:V\rightarrow W$ between EJAs \textbf{positive} when $f(a)\geq 0$ whenever $a\geq 0$. It is \textbf{sub-unital} when $f(1)\leq 1$. We denote the category of EJAs with positive sub-unital maps by $\EJA_{\text{psu}}$.
\end{definition}

\noindent The opposite category $\EJA_{\text{psu}}^\opp$ is an operational effect theory and is remarkably well-behaved (see~\cite{westerbaan2018puremaps} for more details).

\section{Purity}\label{sec:purity}

As discussed in the introduction, the axioms of our reconstruction mostly concern the specific definition of purity from effectus theory. In order to appreciate this definition, let us first take a broader look at the concept of purity.

In quite a variety of topics in quantum information theory, a notion of purity has proved fruitful~\cite{devetak2005distillation,brandao2013resource}. While there is consensus about which states should be considered pure, when talking about pure maps, the situation is more muddled. There are a variety of different definitions in play that each have their benefits and drawbacks.
In this section we will review those different definitions, but first let us consider some properties that would be desirable or expected of an intuitive definition of purity.

First of all, what does it mean to say that a map is `pure'? In a way, saying that a map is pure is saying that it is `fundamental' in some way. 
This can mean multiple things. It could mean that every other map can be made in some way using pure maps, and thus that the pure maps are the basic building blocks of the theory. 
It could also mean that the pure maps are the only transformations that are part of the fundamental theory, other transformations merely reflecting our ignorance of these `true' dynamics. 
For instance, in pure quantum mechanics, the systems are Hilbert spaces, while the only allowed transformations are unitaries. Since this is the fundamental level of the physical theory, all these unitaries can be considered pure. Transitioning to the more general framework of C*-algebras, we also have the liberty to describe classical systems and interactions that do not seem to warrant being called pure, such as the action of throwing away a system by the partial trace map.

We will take the view that a map is pure when it is somehow fundamental to the theory. We will now argue that those maps should form a dagger category.

\begin{definition}
    A \textbf{dagger-category} $C$ is a category equipped with an involutive endofunctor $(\cdot)^\dagger:C\rightarrow C^{\text{op}}$ that is the identity on objects: $A^\dagger = A$. In other words, there is an operation $\dagger$ that sends every map $f:A\rightarrow B$ to some map $f^\dagger: B\rightarrow A$ such that $(f^\dagger)^\dagger = f$, $\id^\dagger = \id$ and $(f\circ g)^\dagger = g^\dagger\circ f^\dagger$.
\end{definition}

Saying that the pure maps should form a dagger category is in fact stating three different things:
\begin{itemize}
    \item \emph{The identity map is pure.} Every physical theory should be able to describe the act of not changing a system.
    \item \emph{The composition of pure maps is pure.} If we describe a fundamental set of transformations, then when two transformations could possibly happen after one another, i.e.\ when they are composable, this combined transformation should also be describable in this fundamental theory and hence be pure.
    \item \emph{The time-reverse of a pure map is pure.} We consider the dagger action as describing the reversal of the arrow of time. Saying that the pure maps have a dagger is then akin to saying that for every fundamental operation, the reversed operationa is also fundamental.
\end{itemize}

If we also wish to describe composite systems, then there is an obvious additional requirement for pure maps that a composite of pure maps should again be pure. In this case the pure maps form a monoidal dagger category\footnote{In a monoidal dagger category it is common to require that the dagger preserves the monoidal structure. We will not need this additional requirement.}.

Now let us go over the definitions of purity found in the literature and see how they compare. 
Probably the most well-known is that of \emph{atomicity} and the related notion of \emph{convex-extremality} used extensively in generalized probabilistic theories~\cite{barrett2007information,chiribella2010probabilistic,chiribella2011informational}. A map $f$ is atomic when any decomposition $f=g_1+g_2$ implies that $g_i = \lambda_i f$. The maps $f$, $g_1$ and $g_2$ here are then required to be `sub-causal'. For causal maps one can consider convex-extremality. A map $f$ is convex extreme when any decomposition $f=\lambda g_1 + (1-\lambda)g_2$ for $0<\lambda<1$ implies that $g_1=g_2=f$. If we take atomicity to be our definition of pure, an immediate problem arises. Consider the C*-algebra $M_n(\C)\oplus M_n(\C)$. The identity can then be written as $\id = \id_1+\id_2$, and hence it is not atomic and thus not pure. Taking convex-extremality as our definition of purity leads to a more subtle problem: the dagger of convex-extreme maps does not have to be sub-unital.

Other definitions of purity are those given by leaks~\cite{selby2018reconstructing}, orthogonal factorizations~\cite{cunningham2017purity} or dilations~\cite{tull2019phdthesis}. Without going into the details, these definitions of purity are in general not closed under a dagger operation. They also need a tensor product in order to be defined. This makes it impossible to define them when no tensor product is available. On a conceptual level there is then also the issue that the purity of a map using these definitions can only be established by considering external systems and hence purity does not seem to be an inherent property of the system and its dynamics.

It should be noted that all these definitions of purity were specifically designed to be applicable to finite-dimensional systems. When considering for instance von Neumann algebras, it is no longer clear that these definitions serve their intended purpose.

\subsection{Filters and compressions}\label{sec:filterscompressions}
Having established that these definitions of purity will not serve for our purposes, we will now introduce the notion of purity arising from effectus theory that we will use for the rest of this paper. This definition basically says that a map is pure when it is a composition of a \emph{filter} and a \emph{compression}. A filter corresponds to doing a measurement while a compression corresponds to forgetting such a measurement took place. Taking these maps as fundamental is then rooted in the view of quantum theory as a description of information processing~\cite{fuchs2002quantum}.

To understand the definition of a compression, let us consider a system $A$ in an effect theory and let $B$ be a `subsystem' of $A$, i.e.\ some system with an embedding map $\pi: B\rightarrow A$. 
If we have some state $\omega:I\rightarrow B$ on this subsystem, then we can view it as a state on the whole system by `forgetting' it was actually defined on the subsystem: $\pi\circ \omega: I\rightarrow A$. Now let $q:A\rightarrow I$ be the effect that `witnesses' whether a state is defined on the subsystem $B$. That is, it is the smallest effect such that $q\circ \pi\circ\omega = 1$ for all states $\omega:I\rightarrow B$. This is the case when $q\circ \pi = 1\circ \pi$. Going in the converse direction, we can for every effect $q$ of $A$ ask which subsystem it witnesses. This will be the subsystem of $A$ where `$q$ is true'. The map that finds the subsystem of $A$ corresponding to $q$ is what we call a compression for $q$.

\begin{definition}\label{def:compression}
    Let $q: A\rightarrow I$ be an effect in an effect theory. We call a system $\{A\lvert q\}$ a \textbf{compression system} for $q$ when there is a map $\pi_q:\{A\lvert q\}\rightarrow A$ such that $1\circ \pi_q = q\circ \pi_q$ that is \textbf{final} with this property: whenever $f:B\rightarrow A$ is such that $1\circ f = q\circ f$ then there is a unique $\cl{f}:B\rightarrow \{A\lvert q\}$ such that the following diagram commutes:
    \[\begin{tikzcd}[ampersand replacement = \&]
    \{A\lvert q\} \arrow{r}{\pi_q}\& A \\
    B \arrow[dotted]{u}{\overline{f}}\arrow{ru}[swap]{f}\&  \\
    \end{tikzcd}\] 
    The map $\pi_q$ is called a \textbf{compression} for $q$\footnote{The term `compression' should not be confused with the maps of the same name in \cite{alfsen2012geometry}. The compressions from that paper correspond to what we will later call `assert maps'.}.
\end{definition}

The universal property of compressions tells us that the system $\{A\lvert q\}$ is the `largest' system such that there is a map $\pi:\{A\lvert q\}\rightarrow A$ with $1\circ\pi = q\circ \pi$, and hence this is the system that we should see as the true subsystem that $q$ witnesses.

\begin{example}
    Let $A=M_n(\C)$ and $B=M_k(\C)$ be complex matrix algebras with $k\leq n$ so that we can view $B$ as a subset of $A$ in the obvious way, i.e.\ when $(c_{ij})$ is matrix in $B$, we can view it as a matrix in $A$ by setting $c_{ij}=0$ when $i,j> k$. Let $\pi:B\rightarrow A$ be this inclusion map, then $\pi$ is a compression for the effect $q\in A$ given by $q=\sum_{i\leq k}\ket{i}\bra{i}$. In fact, it is a compression for any effect $q+p$ where $q$ and $p$ are orthogonal and $\norm{p} < 1$.
\end{example}

\begin{remark}
    It might not be clear from this example why we call such an inclusion map a compression. Originally, compressions were studied in von Neumann algebras~\cite{westerbaan2016universal}. There, unital maps are used. In that case, compressions correspond to certain projection maps. The reason the compression looked like an inclusion in the example is because we used trace-preserving maps instead of unit-preserving maps.
\end{remark}

While a compression can be seen as a map that `forgets' that a state came from a subsystem, a filter is the opposite, describing how a state can be `filtered' to fit inside a subsystem.

\begin{definition}\label{def:filter}
    Let $q:A\rightarrow I$ be an effect in an effect theory. A \textbf{filter} for $q$ is a map $\xi_q:A\rightarrow A_q$ such that $1\circ\xi_q\leq q$ which is \textbf{initial} for this property: for any map $f:A\rightarrow B$ which satisfies $1\circ f\leq q$ there is a unique $\cl{f}:A_q\rightarrow B$ such that the following diagram commutes:
    \[\begin{tikzcd}[ampersand replacement = \&]
    A_q \arrow[dotted,swap]{d}{\cl{f}}\&\arrow[swap]{l}{\xi_q} A\arrow{dl}{f} \\
     B\&  \\
    \end{tikzcd}\]
\end{definition}

\noindent The interpretation of the filter $\xi_q$ is that it represents a non-destructive measurement of $q$ which got a positive outcome, i.e.\ it is a post-selection for $q$. The space $A_q$ is the subsystem where $q$ has a nonzero probability of being true.

\begin{example}
    Again, let $A=M_n(\C)$ and $B=M_k(\C)$ with $k\leq n$ so that we view $B\subseteq A$. Let $q\in A$ be an effect with spectral decomposition $q=\sum_i \lambda_i \ket{v_i}\bra{v_i}$ where the $v_i$ are orthogonal and $\ket{v_i}\bra{v_i}\in B$ and $\lambda_i > 0$. We note that then for any matrix $p\in A$ we get $\sqrt{q}p\sqrt{q}\in B$ (when viewed as a subset). Let $\xi_q: A\rightarrow B$ be defined by $\xi_q(p) = \sqrt{q}p\sqrt{q}$, then $\xi_q$ is a filter for $q$.
\end{example}

\begin{remark}
    In effectus theory, compressions are called \textbf{comprehensions} and filters are called \textbf{quotients}. They arise in a wide variety of settings~\cite{cho2015quotient}. For a full account of the theory of effectuses with quotient and comprehension we refer to Ref.~\cite{basthesis}.
\end{remark}

\begin{remark}
    Consider the deterministic effect $1:A\rightarrow I$. The subsystem associated with $1$ is of course $A$ itself. All the compressions of $1$ are then all the ways in which $A$ can be embedded in itself, i.e.\ they are the isomorphisms of $A$. Isomorphisms of $A$ are also filters for $1$. The filter and compression for the zero effect $0:A\rightarrow I$ is the zero map.
\end{remark}

\begin{remark}
    Due to their universal properties, the compressions and filters of a given effect are unique up to unique isomorphism. To be more specific, if $\pi:\{A\lvert q\}\rightarrow A$ and $f:B\rightarrow A$ are both compressions for $q$, then there is a unique isomorphism $\Theta:B\rightarrow \{A\lvert q\}$ such that $f = \pi\circ\Theta$. Similarly, if $\xi_q:A\rightarrow A_q$ and $g:A\rightarrow B$ are both filters for $q$ then there is a unique isomorphism $\Theta:A_q\rightarrow B$ such that $g=\Theta\circ\xi_q$. 
    Conversely, if we have an isomorphism, then we can compose it in these ways with a filter or compression to create a new filter or compression.
\end{remark}

By these two previous remarks, a filter is more accurately described as a measurement followed by some post-processing in the form of a reversible transformation (and similarly a compression allows some pre-processing).

Using compressions and filters we can define our notion of purity.
\begin{definition}\label{def:pure}
    A map $f:A\rightarrow B$ in an effect theory is \textbf{pure} when $f=\pi\circ\xi$ where $\xi$ is a filter and $\pi$ is a compression.
\end{definition}

As a filter corresponds to measuring and post-processing, and a compression corresponds to forgetting, this definition tells us that a map is pure when it can be written in terms of these fundamental operations: a measurement, reversible post-processing, and finally forgetting some information.

Compared to the previous definitions of purity that were discussed, these pure maps are remarkably well-behaved in a variety of settings:

\begin{proposition}
    Let $\mathbb{E}$ be \textbf{vNA}$_{\text{ncpsu}}$, the category of von Neumann algebras with normal completely positive sub-unital maps or \textbf{EJA}$_{\text{psu}}$, the category of Euclidean Jordan algebras with positive sub-unital maps. Then for every effect in $\mathbb{E}^{\text{op}}$ we can find a compression and a filter and furthermore, the pure maps form a dagger category.
\end{proposition}
\begin{proof}
    The von Neumann algebra case can be found in~\cite{westerbaan2016universal,bramthesis}, while the Jordan algebra case can be found in~\cite{westerbaan2018puremaps}.
\end{proof}

\subsection{Sharpness}

As discussed in the previous section, a compression ${\pi_q: \{A\lvert q\}\rightarrow A}$ tells us that $\{A\lvert q\}$ is a subsystem of $A$. We also know that $1\circ\pi_q = q\circ\pi_q$, but $q$ might not be the smallest effect with this property. This is because an effect can be \emph{fuzzy}, meaning that it does not make a sharp distinction between where it holds true, and where it doesn't. In contrast, if $q$ \emph{is} the smallest effect with this property, then we call it \emph{sharp}, since it sharply delineates its subspace.

\begin{definition}
    Let $f:A\rightarrow B$ be a transformation in an effect theory. The \textbf{image} of $f$, when it exists, is the smallest effect $q:B\rightarrow I$ such that $q\circ f = 1\circ f$, i.e.\ if $p:B\rightarrow I$ is also such that $p\circ f = 1\circ f$, then $q\leq p$. We denote the image of $f$ by $\im{f}$.
\end{definition}

By the discussion above, we would call an effect sharp when $\im{\pi_q}=q$. Instead we will use the slightly more general formulation used in~\cite{basthesis} that turns out to be equivalent (see Proposition~\ref{prop:floorceiling}):

\begin{definition}
    Let $q:A\rightarrow I$ be an effect. We call $q$ \textbf{sharp} when there is some transformation $f:B\rightarrow A$ such that $\im{f} = q$.
\end{definition}

Unfortunately, it seems there is little we can say about sharp effects in a general effect theory, without specifying additional assumptions. Suppose $q$ is a sharp effect. As discussed, this means that there is some subsystem where $q$ is `true'. Similarly, we would expect there also to be some subsystem where $q$ is `false', i.e.\ we would expect its negation $q^\perp$ to be sharp as well. This is our first assumption.

The second set of assumptions relies on the interplay between filters and compressions of sharp effects. Let $\pi_q:\{A\lvert q\}\rightarrow A$ be a compression for a sharp effect $q$. Let $\omega:I\rightarrow \{A\lvert q\}$ be a state. 
As the state is part of this subsystem, it satisfies the effect $q$ with certainty. 
The state $\pi_q\circ\omega$ is then the same state on $A$ where we have forgotten that $q$ holds. As $\pi_q$ is a pure map, we can consider its adjoint, i.e.\ the time-reverse, $\pi_q^\dagger$. As the compression $\pi_q$ forgets that the effect $q$ holds for $\omega$, $\pi_q^\dagger$ should `remember it', i.e.\ it is a post-selection for $q$, and so it is a filter for $q$. But a post-selection after we already knew that the effect holds is just doing nothing, and hence we should have $\pi_q^\dagger \circ \pi_q = \id$. These are our last assumptions. In the parlance of dagger theories: compressions for sharp effects should be isometries, and compressions and filters for sharp effects are adjoint.

\subsection{Pure Effect Theories}

As we now have illustrated all the necessary assumptions, let us define the main object we will study in this paper.

\begin{definition}\label{def:PET}
    A (monoidal) \textbf{Pure Effect Theory} (PET) is a (monoidal) effect theory satisfying the following properties.
    \begin{enumerate}[label=({P}\theenumi), ref=P\theenumi]
        \item \label{pet:filtcompr} All effects have filters and compressions.
        \item \label{pet:dagger} The pure maps form a (monoidal) dagger-category.
        \item \label{pet:images} All maps have images.
        \item \label{pet:sharpnegation} The negation of a sharp effect is sharp (if $q$ is sharp, then $q^\perp$ is sharp)
        \item \label{pet:sharpadjoint} Filters of sharp effects are adjoint to its compressions (if $\pi_q$ is a compression for sharp $q$, then $\pi_q^\dagger$ is a filter for $q$, and vice versa).
        \item \label{pet:sharpisometry} Compressions of sharp effects are isometries ($\pi_q^\dagger\circ\pi_q = \id$ for sharp $q$).
    \end{enumerate}
\end{definition}

\begin{remark}
    All these properties are taken from~\cite{basthesis}. They are a subset of what is required of a \textbf{$\dagger$-effectus}. An effectus satisfying points 1,3 and 4 is called a \textbf{$\diamond$-effectus}. The only known examples of $\dagger$-effectuses are $\textbf{vNA}^\opp_{\text{cpsu}}$~\cite{bramthesis} and $\textbf{EJA}^\opp_{\text{psu}}$~\cite{westerbaan2018puremaps}.
\end{remark}

\begin{remark}
    Several of these properties are closely related to more familiar categorical definitions. It is shown in Ref~\cite{basthesis} that an effect theory has all compressions if and only if it has all \textbf{kernels} (a compression of $q$ is a kernel of $q^\perp$). It has all \textbf{cokernels} if and only if all maps have an image and every sharp effect has a filter (a filter is the cokernel of a compression). The assumptions~\ref{pet:sharpadjoint} and \ref{pet:sharpisometry} can equivalently be stated as ``all kernels are dagger-kernels, and the dagger of a kernel is a cokernel'', and hence the subcategory of pure maps of a PET is a \textbf{dagger kernel category}~\cite{heunen2010quantum}.
\end{remark}

\noindent The definition of a PET allows us to formally state our main results:
\begin{theorem*}[\textbf{\ref{theor:OETEJA}}]
    Let $\mathbb{E}$ be an operational pure effect theory. Then there exists a functor into the opposite category of Euclidean Jordan algebras and sub-unital positive maps $F:\mathbb{E}\rightarrow \EJA_{\text{psu}}^{\text{op}}$ such that $[0,1]_{F(A)} \cong \text{Eff}(A)$. Furthermore, this functor is faithful if and only if $\mathbb{E}$ satisfies local tomography.
\end{theorem*}

\begin{theorem*}[\textbf{\ref{theor:compositealgebras}}]
    Let $\mathbb{E}$ be a monoidal operational pure effect theory. Then the above functor restricts either to the category of real C*-algebras or to the category of complex C*-algebras. 
\end{theorem*}

In this last theorem we expect that $F$ is faithful if and only if $\mathbb{E}$ satisfies tomography, but showing this requires establishing that $F$ is monoidal, which is currently an open question. For more discussion regarding this we refer to Section~\ref{sec:quantum}.

\section{Properties of PETs}\label{sec:firsttwo}
In this section we will discuss the basic properties of effect theories, compressions and filters. Let us start with a few results that we will use without further reference:
\begin{proposition} The following are true in any effect theory.
    \begin{enumerate}
        \item Let $p,q \in \pred(A)$ be effects such that $p\leq q$, and let $f:B\rightarrow A$ be some map, then $p\circ f \leq q\circ f$.
        \item Let $\Theta: A\rightarrow B$ be an isomorphism, i.e.\ a map that has a 2-sided inverse $\Theta^{-1}:B\rightarrow A$, then $1\circ \Theta = 1$.
    \end{enumerate}
\begin{proof}~
    \begin{enumerate}
    \item By definition, $p\leq q$ if and only if there is some $r\in \pred(A)$ such that $p+r = q$. Maps in an effect theory preserve addition and hence $q\circ f = (p+r)\circ f = p\circ f + r\circ f$ so that indeed $p\circ f \leq q\circ f$.
    \item Let $1\circ \Theta = p$. We need to show that $p^\perp = 0$. Of course $p\circ \Theta^{-1} = 1$ and hence $1\circ \Theta^{-1} = (p+p^\perp)\circ \Theta^{-1} = p\circ \Theta^{-1} + p^\perp \circ \Theta^{-1} = 1 + p^\perp \circ \Theta^{-1}$. Since the only effect summable with $1$ is $0$, we get $p^\perp\circ \Theta^{-1} = 0$. As a result $0 = 0\circ \Theta = p^\perp\circ\Theta^{-1}\circ \Theta = p^\perp$ and we are done. \qedhere
    \end{enumerate}
\end{proof}
    
\end{proposition}

The following proposition contains a few basics results regarding compressions and filters that can be found in Refs.~\cite{cho2015introduction,basthesis}.

\begin{proposition} \label{prop:quotcompr}
	\cite{cho2015introduction} Let $q:A\rightarrow I$ be an effect and let $\pi_q:\{A\lvert q\}\rightarrow A$ be a compression for $q$ and $\xi_q:A\rightarrow A_q$ a filter. 
	\begin{itemize}
		\item Let $\Theta:B\rightarrow \{A\lvert q\}$ be an isomorphism, then $\pi\circ \Theta$ is another compression for $q$. Conversely, for any two final compressions $\pi$ and $\pi^\prime$ for $q$ there is a isomorphism $\Theta$ such that $\pi^\prime = \pi\circ \Theta$.
		\item Let $\Theta:A_q\rightarrow B$ be an isomorphism, then $\Theta\circ \xi$ is another filter for $q$. Conversely for any two initial filters $\xi$ and $\xi^\prime$ for $q$ there exists a isomorphism $\Theta$ such that $\xi^\prime = \Theta\circ \xi$.
		\item Isomorphisms are filters and compressions for the truth effect $1$.
		\item Zero-maps are filters and compressions for the falsity effect $0$.
		\item As an effect algebra, the set of effects of $A_q$ is isomorphic to the \textbf{downset} of $q$: Eff$(A_q)\cong \{p \in \text{Eff}(A)~;~ p\leq q\}$.
	\end{itemize}
\end{proposition}

Note that all the properties above are relatively straightforward to prove using the universal properties of filters and compressions, except for the last point which takes a bit more work. This last point will be important when we will show the equivalence of atomic and pure effects in Proposition \ref{prop:atomicstate}.

\begin{definition}
	We call a transformation $f:A\rightarrow B$ in an effect theory \textbf{unital} when $1\circ f = 1$. We call it \textbf{faithful} when $q\circ f = 0$ implies that $q=0$. This is true if and only if $f$ has an image and $\im{f}=1$.
\end{definition}

\begin{proposition}\label{prop:faithfulfilters}
	\cite{cho2015introduction,basthesis} In an effect theory with images, filters and compressions the following are true.
	\begin{itemize}
		\item Compressions are unital.
		\item Filters are faithful.
        \item Let $\xi_q$ be a filter for $q$, then $1\circ \xi_q = q$.
	\end{itemize}
\end{proposition}

\begin{definition}
Let $q$ be an effect and let $\pi$ be a compression for $q$. The \textbf{floor} of $q$ is defined as $\floor{q}:= \im{\pi}$. The \textbf{ceiling} is defined as the De Morgan dual: $\ceil{p} = \floor{p^\perp}^\perp$.
\end{definition}

The ceiling of an effect corresponds to what is called the \emph{face} of an effect in \cite{chiribella2011informational}. In quantum theory, the ceiling of an effect is the projection onto the range of the effect. The floor is the projection onto the space where the effect acts as the identity.

\begin{proposition}\label{prop:floorceiling}
	\cite{basthesis} In an effect theory with images and compressions, the following are true for any effect $q$.
	\begin{itemize}
		\item $\floor{q}\leq q$.
		\item $\floor{\floor{q}}=\floor{q}$.
		\item $p\leq q\implies \floor{p}\leq \floor{q}$.
		\item $\ceil{q}\circ f = 0 \iff q\circ f = 0$.
		\item $\ceil{q\circ f} = \ceil{\ceil{q}\circ f}$.
		\item $q$ is sharp if and only if $\floor{q}=q$.
	\end{itemize}
\end{proposition}

\noindent Note that this proposition implies that the floor of $q$ is the largest sharp predicate below $q$ so that the name is indeed well-chosen.

\begin{proposition}\label{prop:sharplattice}
	\cite[208IX]{basthesis} In an effect theory with images, compressions, and filters additionally satisfying \ref{pet:sharpnegation}, the sharp effects form a lattice.
\end{proposition}

\subsection{Assert maps}\label{sec:assertmaps}

Let $p$ be a sharp effect and let $\pi_p$ be a compression for it. By \ref{pet:sharpadjoint} the map $\pi_p^\dagger$ is a filter for $p$ and by \ref{pet:sharpisometry} we have $\pi_p^\dagger\circ \pi_p = \id$.

\begin{definition}
	Let $p:A\rightarrow I$ be a sharp effect in a PET and let $\pi_p: \{A\lvert p\}\rightarrow A$ be a compression for it. We define the \textbf{assert map} $\asrt_p:A\rightarrow A$ for $p$ by $\asrt_p = \pi_p\circ \pi_p^\dagger$. 
\end{definition}

\begin{remark}
    The definition of the assert map does not depend on the choice of compression. If $\pi_p^\prime$ is also a compression for $p$ then $\pi_p^\prime = \pi_p\circ\Theta_1$ for some isomorphism $\Theta_1$ by Proposition~\ref{prop:quotcompr}, and similarly since $(\pi_p^\prime)^\dagger$ is a filter by \ref{pet:sharpadjoint} we have $(\pi_p^\prime)^\dagger = \Theta_2\circ\pi_p^\dagger$ for some other isomorphism $\Theta_2$. Now $\id = (\pi_p^\prime)^\dagger\circ \pi_p^\prime = \Theta_2 \circ \pi_p^\dagger \circ \pi_p \circ \Theta_1 = \Theta_2\circ \Theta_1$ and hence $\Theta_2 = \Theta_1^{-1}$. As a result $\pi_p^\prime \circ (\pi_p^\prime)^\dagger = \pi_p\circ \Theta_1\circ \Theta_1^{-1} \circ \pi_p^\dagger = \pi_p \circ \pi_p^\dagger = \asrt_p$.
\end{remark}
\begin{remark}
    Since $\xi_p^\dagger$ is also a compression for $p$ we could also define the assert map as $\asrt_p = \xi_p^\dagger \circ \xi_p$ for any filter of $p$.
\end{remark}


\begin{proposition} \label{prop:assertmaps}
	\cite{basthesis} Let $p$ be a sharp effect and let $f$ be any composable map in a PET. The following are true:
	\begin{enumerate}
		\item $\asrt_p\circ \asrt_p = \asrt_p$.
		\item $\im{\asrt_p} = p$.
		\item $1\circ \asrt_p = p$.
		\item $\im{f}\leq p \iff \asrt_p\circ f = f$.
		\item $1\circ f \leq p \iff f\circ \asrt_p = f$.
	\end{enumerate}
\end{proposition}
\begin{proof} ~
	\begin{enumerate}
		\item $\asrt_p\circ \asrt_p = \pi_p\circ\pi_p^\dagger\circ\pi_p\circ\pi_p^\dagger = \pi_p\circ \id\circ \pi_p^\dagger = \asrt_p$.

		\item Suppose $q\leq \im{\asrt_p}^\perp$, then $0 = q\circ \asrt_p = q\circ \pi_p\circ \pi_p^\dagger$. Because $\pi_p^\dagger$ is a filter, it is faithful by Proposition~\ref{prop:faithfulfilters}. As a result $q\circ \pi_p =0$ so that $q\leq \im{\pi_p}^\perp = p^\perp$. Taking $q=\im{\asrt_p}^\perp$ we then have $p \leq \im{\asrt_p}$. For the other direction we have $p\circ \asrt_p = p\circ \pi_p \circ \pi_p^\dagger = 1\circ \pi_p\circ \pi_p^\dagger = 1\circ \asrt_p$ so that $p\geq \im{\asrt_p}$.

		\item $1\circ \asrt_p = 1\circ \pi_p \circ \pi_p^\dagger = 1\circ \pi_p^\dagger = p$ by the unitality of $\pi_p$ (Proposition~\ref{prop:faithfulfilters}).

		\item If $\asrt_p\circ f = f$, then since $1\circ \asrt_p\circ f = p\circ \asrt_p\circ f$ we get $\im{f} = {\im{\asrt_p\circ f}} \leq \im{\asrt_p} = p$. For the other direction, if $\im{f}\leq p$, then $p\circ f = 1\circ f$ so that by the universal property of compressions $f = \pi_p\circ \cl{f}$ for some $\cl{f}$. Now $\cl{f} = \id\circ \cl{f} = \pi_p^\dagger\circ \pi_p \circ \cl{f} = \pi_p^\dagger \circ f$ so that $f = \pi_p \circ \cl{f} = \pi_p\circ \pi_p^\dagger\circ f = \asrt_p\circ f$.
		\item Suppose $f\circ \asrt_p = f$, then $1\circ f = (1\circ f)\circ \asrt_p \leq 1\circ \asrt_p = p$. For the other direction, if $1\circ f \leq p$, then by the universal property of filters we get $f = \cl{f}\circ \pi_p^\dagger$ for some $\cl{f}$. Now $\cl{f} = \cl{f}\circ \id = \cl{f} \circ \pi_p^\dagger \circ \pi_p = f \circ \pi_p$ so that $f = \cl{f}\circ \pi_p^\dagger = f\circ \pi_p\circ \pi_p^\dagger = f\circ \asrt_p$. \qedhere
	\end{enumerate}
\end{proof}

\noindent Recall that in an effect algebra the addition operation is a partial operation. This will allow us to talk about orthogonality.
\begin{definition}
	Let $p,q\in$ Eff$(A)$ be sharp effects. We call them \textbf{orthogonal} and write $p\perp q$ when $p$ and $q$ are summable. That is, when $p+q$ is defined and therefore $p+q\leq 1$. We call two arbitrary effects (not necessarily sharp) orthogonal when their ceilings are orthogonal: $p\perp q \iff \ceil{p}\perp \ceil{q}$.
\end{definition}
Note that for non-sharp effects, being summable is weaker than being orthogonal, as for instance $\frac12 p$ is always summable with itself (assuming we have a scalar acting like $\frac12$).
\begin{proposition}\label{prop:orthog}
	Sharp effects $p,q\in\pred(A)$ are orthogonal if and only if $p\leq q^\perp \iff q\leq p^\perp$. In a PET $p\perp q$ if and only if $q\circ \asrt_p = 0$.
\end{proposition}
\begin{proof}
	Suppose $p+q$ exists. Then there is a $r=(p+q)^\perp$ such that $p+q+r = 1$, but then by uniqueness of the complement we get $q+r=p^\perp$ so that by definition $q\leq p^\perp$ (and this argument is obviously symmetric in $p$ and $q$). For the other direction, when $q\leq p^\perp$ there is an $r$ such that $q+r=p^\perp$, but then $q+r+p = p^\perp+p = 1$. But then $r^\perp = q+p$ so that $q$ and $p$ are indeed summable.

	Now suppose we have assert maps. We have $\im{\asrt_p} = p$, so that for any $q\leq p^\perp =\im{\asrt_p}^\perp$ we get $q\circ \asrt_p \leq \im{\asrt_p}^\perp\circ \asrt_p = 0$. For the other direction if we have $q\circ \asrt_p = 0$ then by definition of the image we have $q\leq \im{\asrt_p}^\perp =p^\perp$.
\end{proof}



\begin{proposition}\label{prop:sharpadd}
	Let $p,q\in \pred(A)$ be effects in a PET.
	\begin{enumerate}
		\item Let $p,q$ be sharp. If $p\perp q$ then $p+q$ is sharp and is the least upper bound of $p$ and $q$.
		\item Suppose $p+q$ is defined and that $p$ and $p+q$ are sharp, then $q$ is sharp.
		\item Let $p,q$ be sharp with $p\leq q$, then $q-p$ is sharp.
	\end{enumerate}
\end{proposition}
\begin{proof}~
\begin{enumerate}
	\item Since $p\perp q$, $p+q$ is again an effect and it is obviously an upper bound of $p$ and $q$. We know that there is a least upper bound\footnote{There is a subtlety here where the least upper bound in the set of sharp effects might not be the least upper bound in the set of all effects, but in fact, if $p,q\leq a$ then also $p,q\leq \floor{a}$ when $p$ and $q$ are sharp so that the least upper bound in the set of all effects must be sharp, and hence must coincide with the least upper bound in the set of sharp effects.} $p\vee q$ by Proposition~\ref{prop:sharplattice}. Now since $p\vee q\geq p,q$ we get $p\circ\asrt_{p\vee q}=p$ and $q\circ\asrt_{p\vee q} = q$ by Proposition~\ref{prop:assertmaps} so that $(p+q)\circ\asrt_{p\vee q} = p+q$. Again by Proposition~\ref{prop:assertmaps} $p+q = 1\circ(p+q)\leq p\vee q$ so that we must have $p+q = p\vee q$. Since the right-hand side is sharp we are done.
	\item $p+q$ is sharp so by \ref{pet:sharpnegation} $(p+q)^\perp$ is sharp. Since $p\leq p+q$ we have $p^\perp\geq (p+q)^\perp$ so that $p\perp (p+q)^\perp$. By the previous point $p+(p+q)^\perp$ is then sharp. Observing that $q+ (p+(p+q)^\perp) = 1$ we get $q^\perp = p+(p+q)^\perp$ by the uniqueness of the complement so that $q^\perp$ is sharp. Again by \ref{pet:sharpnegation} we see that $q$ is sharp.
	\item By the previous point $p+ (q-p) = q$ is sharp so that $q-p$ must be sharp as well. \qedhere
\end{enumerate}
\end{proof}

\noindent There is a lot more that can be done in this abstract situation of effect theories with filters and compressions. For the interested reader we refer to Refs.~\cite{cho2015introduction,basthesis}. We will now switch to a more concrete setting.

\section{From operational PETs to Jordan algebras}\label{sec:opeffect}

In this section we will study PETs in a more familiar convex setting by working with operational PETs. The goal is to show that in this setting, the systems correspond to Euclidean Jordan algebras, and hence most of the structure of quantum theory is recovered. 
We will first derive a diagonalisation theorem in Section~\ref{sec:diagonalisation}. Then, in Section~\ref{sec:duality} we will construct an inner product on the effect spaces. In combination with some further technical arguments presented in Appendix~\ref{sec:proofofeja}, this will show that the systems correspond to EJAs.

For the duration of this section we will assume that we have fixed some operational PET, and that $A$ and $B$ are systems therein. Furthermore $V$ and $W$ will denote the order unit spaces associated to them: $\pred(A) \cong [0,1]_V$. Sharpness, ceilings and floors in $V$ and $W$ are all defined by the same notion on effects in $A$ and $B$.

\subsection{Diagonalisation}\label{sec:diagonalisation}
We will construct a diagonalisation in terms of sharp effects using properties of the ceiling of effects. That is: we will show that for any $v\in V$ we can find a collection of eigenvalues $\lambda_i\in \R$ and a set of sharp effects $p_i\in V$ that are all orthogonal to each other such that $v=\sum_i \lambda_i p_i$. The following proposition collects the needed properties of the ceiling established in Section~\ref{sec:firsttwo}:
\begin{proposition}~\label{prop:ceilings}
	Let $p,q\in \pred(A)$  and $f:B\rightarrow A$.
	\begin{itemize}
		\item If $p\leq q$ then $\ceil{p}\leq \ceil{q}$ and $\floor{p}\leq \floor{q}$.
		\item Let $c>0$ be a scalar. Then $\ceil{cp}=\ceil{p}$.
		\item $p\circ f = 0$ iff $\ceil{p}\circ f = 0$.
	\end{itemize}
\end{proposition}
\begin{proof}
	All the points are straight from Proposition~\ref{prop:floorceiling} except for $c>0$ implying that $\ceil{cp}=\ceil{p}$. This follows from the point in Proposition~\ref{prop:floorceiling} that $\ceil{q\circ f}=\ceil{\ceil{q}\circ f}$ by letting $f=p$ and $q=c$ and observing that in our setting $\ceil{c}=1$ when the scalar $c$ is greater than zero.
\end{proof}

An order unit space has a norm given by $\norm{v}:=\inf\{r>0~;~-r1\leq v\leq r1\}$, and this norm induces a topology. Let $C := \{v\in V~;~ v\geq 0\}$ be the positive cone of $V$. Its interior with respect to the norm-topology is denoted by $C^\circ$. It is straightforward to see that $v\in C^\circ$ if and only if there is some $\epsilon>0$ such that $\epsilon 1 \leq v$.

\begin{lemma} \label{lem:var}
	Let $C^\circ$ denote the interior of the positive cone of $V$ and $\partial C = C\backslash C^\circ$ denote the boundary. Let $v\in C$, then 
		\begin{enumerate}
			\item $\norm{v}1 - v \in \partial C$.
			\item If $\norm{v}<1$, then $1-v=v^\perp \in C^\circ$.
			\item If $v$ is sharp, we have $\norm{v}<1 \implies v=0$.
			\item If $\norm{v}<1$ and $v\leq p$ where $p$ is sharp, then $\ceil{p-v}=p$.
            \item If $\norm{v}<1$ and $\ceil{v}\perp p$ where $p$ is sharp, then $\floor{p+v} = p$.
		\end{enumerate}
\end{lemma}
\begin{proof} ~
	\begin{enumerate}
	\item As $v\leq \norm{v}1$, if $\norm{v}1-v \not\in\partial C$ then $v-\norm{v}1\in C^\circ$ so that there must be an $\epsilon>0$ such that $\epsilon 1 \leq v-\norm{v}1$ which means that $v-(\norm{v}+\epsilon)1 \geq 0$ contradicting the defining property of the norm. 
	\item Of course $0\leq v\leq \norm{v}1$ and hence $0\leq 1-\norm{v}1 \leq 1-v=v^\perp$. Since $1-\norm{v}1>0$ we conclude that $v^\perp \in C^\circ$.
	\item Let $v$ be sharp with $\norm{v}<1$. Then by the previous point $v^\perp \in C^\circ$ so that $\epsilon 1\leq v^\perp$ for some $\epsilon>0$. Then $1=\ceil{1}=\ceil{\epsilon 1} \leq \ceil{v^\perp} = v^\perp$ because $v^\perp$ is sharp. But then $v^\perp = 1$ so indeed $v=0$.
	\item Let $v\leq p$ with $\norm{v}<1$, and denote $q:=\ceil{p-v}\leq p$. Then we can write $p=q+r$ where $r=p-q$ is a sharp effect by Proposition~\ref{prop:sharpadd}. Now $p-v\leq \ceil{p-v}=q$ so that $r=p-q\leq v$. Taking the norm on both sides gives $\norm{r} \leq \norm{v}<1$ so that by the previous point $r=0$. So indeed $p=q+0 = \ceil{p-v}$.
    \item Note first that because $\ceil{v}\perp p$ and hence $\ceil{v}+p\leq 1$ we also have $v+p\leq 1$. We calculate $\floor{p+v} = \ceil{(p+v)^\perp}^\perp = \ceil{1-p-v}^\perp = \ceil{p^\perp - v}^\perp$. Since $\ceil{v}\perp p$ we have $v\leq \ceil{v}\leq p^\perp$ and since $\norm{v}<1$ the previous point applies and we have $\ceil{p^\perp - v} = \ceil{p^\perp}$ so that indeed $\floor{p+v} = \ceil{p^\perp -v}^\perp = \ceil{p^\perp}^\perp = \floor{p}=p$. \qedhere
	\end{enumerate}
\end{proof}
We need to show that the order unit spaces we deal with are `finite-rank' in a suitable way. This turns out to follow from the next lemma.

\begin{lemma}\label{lem:finitedimrank}
    Let $0\leq v\leq 1$ in $V$. We have $v\in C^\circ \iff \ceil{v}=1$.
\end{lemma}
\begin{proof}
    First of all, when $v\in C^\circ$ we have $\epsilon 1\leq v$ and hence $\ceil{v}\geq \ceil{\epsilon 1} = 1$, so let us go in the opposite direction.

    Let $0\leq v\leq 1$ and suppose $\ceil{v}=1$. We want to show that there exists an $\epsilon>0$ such that $\epsilon 1\leq v$. Let $\omega$ be a unital state (so that $1\circ \omega = 1$) and suppose $v\circ \omega = 0$, then $\ceil{v}\circ\omega = 1\circ\omega = 0$ by the last point of Proposition~\ref{prop:ceilings}. This is of course a contradiction so that $v\circ \omega>0$ for all unital states. Define $f:$ St$_1(A)\rightarrow [0,1]$ by $f(\omega)=v\circ \omega$ where St$_1(A)$ denotes the set of unital states on $A$. By assumption St$_1(A)\subset V_A^*$ is closed. It is also bounded, and since $V_A^*$ is a finite-dimensional vector space we conclude that St$_1(A)$ is compact. Therefore the image of $f$ will be some compact subset of $[0,1]$. In particular there is an $\omega$ that achieves the minimum, which has to be strictly bigger than zero: $v\circ\omega \geq \epsilon$. By order-separation we then get $v\geq \epsilon 1$ which shows that $v\in C^\circ$.
\end{proof}

\begin{remark}
    This lemma very explicitly requires the state-space to be closed, and $V$ to be finite-dimensional. Many of the following results can be proved without using finite-dimensionality or the closure of state-space if one adds the truth of this lemma to the list of assumptions on our order unit space.
\end{remark}

\begin{proposition} \label{prop:ceilint}
    Let $v\in [0,1]_V$ with $\norm{v}=1$, then $\floor{v}\neq 0$.
\end{proposition}
\begin{proof}
	By Lemma~\ref{lem:var} $\norm{v}1-v \not\in C^\circ$ and hence by Lemma~\ref{lem:finitedimrank} $\ceil{\norm{v}1-v} \neq 1$. Supposing now that $\norm{v}=1$ we then immediately get $\floor{v}^\perp = \ceil{v^\perp} = \ceil{1-v} = \ceil{\norm{v}1 -v} \neq 1$ and hence $\floor{v}\neq 0$.
\end{proof}



\begin{lemma}\label{lem:orthindep}
	Let $\{p_i\}$ be a finite set of non-zero orthogonal sharp effects (where orthogonal refers to the fact that $p_i+p_j\leq 1$ if $i\neq j$). Then they are linearly independent.
\end{lemma}
\begin{proof}
	Reasoning towards contradiction, assume that there is a non-trivial linear combination of the orthogonal sharp effects. Then without loss of generality $p_1 = \sum_{i>1} \lambda_i p_i$. Since all the $p_i$ are orthogonal we note that $p_j\circ \asrt_{p_i} = 0$ when $i\neq j$ by Proposition~\ref{prop:orthog}, so that $p_1 = p_1\circ\asrt_{p_1} = \sum_{i>1}\lambda_i p_i\asrt_{p_1} = 0$, a contradiction.
\end{proof}

\noindent We can now prove our diagonalisation theorem.

\begin{theorem}\label{theor:diag}
	Let $v\in[0,1]_V$. There is a $k\in \N$ and a strictly decreasing sequence of scalars $\lambda_1>\ldots>\lambda_k>0$ such that $v=\sum_{i=1}^k \lambda_i p_i$ where the $p_i$ are non-zero orthogonal sharp effects.
\end{theorem}
\begin{proof}
If $v=0$ the result is trivial, so assume that $v\neq 0$. 
Let $v^\prime = \norm{v}^{-1} v$ so that $\norm{v^\prime} = 1$. Write $v^\prime = \floor{v^\prime} + w$ for $w=v^\prime - \floor{v^\prime}$. By Proposition~\ref{prop:ceilint} $\floor{v^\prime} \neq 0$. 
Note that $\norm{w}\neq 1$ since otherwise we would get a non-zero $\floor{w}$ so that $\floor{w}+\floor{v^\prime}\leq v^\prime$ implying that $\floor{\floor{w}+\floor{v^\prime}} = \floor{w}+\floor{v^\prime}\leq \floor{v^\prime}$ by Proposition~\ref{prop:sharpadd}. 
Now since $w=v^\prime-\floor{v^\prime}\leq 1 - \floor{v^\prime}$ we can write $\ceil{w}\leq \ceil{1-\floor{v^\prime}} = \ceil{\floor{v^\prime}^\perp} = \floor{\floor{v^\prime}}^\perp = \floor{v^\prime}^\perp$ so that $\ceil{w} \perp \floor{v^\prime}$.

Now using $v = \norm{v}v^\prime$ we see that we can write $v = \lambda_1 p + w$ where $p$ is a sharp effect, $\lambda_1 = \norm{v}$ and $w$ is an effect orthogonal to $p$ with $\norm{w}<\norm{v}$. We can now repeat this procedure for $w$. This has to end at some point for if it would not then we get an infinite sequence of orthogonal sharp effects $\{p_i\}_{i=1}^\infty$ which by Lemma~\ref{lem:orthindep} can only be the case when the space is infinite-dimensional.
\end{proof}

\noindent This diagonalisation is unique in the following sense:

\begin{proposition}
	Let $v = \sum_{i=1}^k \lambda_i p_i$ and $v=\sum_{j=1}^l \mu_j q_j$ where $\lambda_i>\lambda_j>0$ and $\mu_i>\mu_j>0$ for $i<j$ and all the $p_i$ and $q_i$ are sharp and non-zero with all the $p_i$ being orthogonal and all the $q_i$ being orthogonal. Then $k=l$, $\lambda_i=\mu_i$ and $p_i=q_i$ for all $i$.
\end{proposition}
\begin{proof}
	Note first that if $v=0$ that then $k=l=0$ and hence we are done. So assume that $v\neq 0$. It is clear that $\norm{v}=\lambda_1=\mu_1$. Consider $v^\prime = \lambda_1^{-1}v$. Now $\floor{v^\prime}=\floor{p_1 + \sum_{i>1}\lambda_1^{-1}\lambda_i p_i} = p_1$ by application of the last point of Lemma~\ref{lem:var}. Using the second decomposition we also get $\floor{v^\prime} = q_1$, so that $p_1=q_1$. Now we can consider $v_2 = v-\lambda_1p_1$ and continue the procedure.
\end{proof}

We can do this diagonalisation for arbitrary positive elements, because we know that for any positive $a$, $0\leq \norm{a}^{-1}a\leq 1$, so that we can diagonalise $\norm{a}^{-1}a$ and then rescale. Now for an arbitrary element $a$ (not necessarily positive) we have $-n1\leq a \leq n1$ for some $n$, so that $0\leq a+ n1\leq 2n1$. This gives us a diagonalisation $a+n1 = \sum_i\lambda_i p_i$, so that $a = \sum_i\lambda_i p_i - n1 = \sum_i(\lambda_i-n)p_i + n(1-\sum_i p_i)$. As a corollary:
\begin{proposition}
	Any vector $v\in V$ can be written as $v= v^+ - v^-$ where $v^+, v^-\geq 0$ are orthogonal.
\end{proposition}

We can get a more finegrained diagonalisation than just the one in terms of sharp effects.

\begin{definition}
	We call a non-zero sharp effect $p$ \textbf{atomic} when for all $q$ with $0\leq q\leq p$ we have $q=\lambda p$. Or in other words when $\downarrow p \cong [0,1]$ where $\downarrow p$ denotes the \textbf{downset} of $p$.
\end{definition}

\begin{proposition}\label{prop:sharprep}
	Each non-zero sharp effect can be written as a sum of atomic effects.
\end{proposition}
\begin{proof}
	Let $p$ be a non-zero sharp effect. If $p$ is atomic we are done, so suppose it is not. Then we can find an $0\neq a< p$ such that $a$ is not a multiple of $p$. By the diagonalisation theorem $a=\sum_i \lambda_i q_i$ where the $q_i$ are non-zero sharp effects and $0<\lambda_i \leq 1$. At least one of these $q_i$ is not equal to $p$ since otherwise $a$ would be a multiple of $p$. Pick this $q_i$, then obviously $\lambda_i q_i\leq a \leq p$ and by taking ceilings $q_i=\ceil{q_i}=\ceil{\lambda_i q_i}\leq \ceil{p}=p$. Since we know that $q_i\neq p$ we must have $q_i<p$ so that we can write $p=q_i + (p-q_i)$, a sum of two orthogonal sharp effects. We can repeat the process for $q_i$ and for $p-q_i$ which are both smaller sharp effects. By finite-dimensionality and Lemma~\ref{lem:orthindep} we see that this process has to end at some point, which means at the final step we are only left with atomic effects.
\end{proof}

\begin{corollary}\label{cor:atomicspectrum}
    Any $v\in V$ can be written as $v=\sum_i \lambda_i p_i$ where $\lambda_i\in \R$ and the $p_i$ are atomic and orthogonal.
\end{corollary}
\begin{proof}
    First diagonalise $v$ in terms of sharp effects, and then write every sharp effect as a sum of atomic effects.
\end{proof}



\subsection{Duality}\label{sec:duality}
In this section we will study pure states and effects and show that they are related trough an interesting duality.

The following proposition establishes that our definition of purity coincides with atomicity when considering states and effects. This correspondence does not hold for arbitrary maps, as there are pure maps that aren't atomic.

\begin{proposition} \label{prop:purestate}
	An effect $q:A\rightarrow I$ is pure if and only if its corresponding effect $q\in [0,1]_V$ is proportional to an atomic effect. A unital state is pure if and only if its image is atomic.
\end{proposition}
\begin{proof}
    When $q=0$ this is trivial, so assume that $q\neq 0$.

	Suppose $q:A\rightarrow I$ is pure, then it can be written as $q=\pi_s\circ \xi_t$ where $\xi_t: A\rightarrow B$ and $\pi_s:B\rightarrow I$. 
    The image $\im{q}$ is a scalar and must be sharp. The only sharp scalars are 0 and 1, so when $q\neq 0$ we must have $\im{q}=1$ which gives $1=\im{\pi_s\circ\xi_{t}}\leq \im{\pi_s} = s$ so that $s=1$. Compressions for the unit effect are isomorphisms, and hence $\pi_1:B\rightarrow I$ is an isomorphism so that $\pi_1\circ \xi_t = \xi_t^\prime$ is also a filter and $q=\xi_t^\prime$. 
    Of course $q=1\circ q = 1\circ \xi_t^\prime = t$ so that $t=q$. 
    The filter has type $\xi_t^\prime:A\rightarrow I$ so that $I\cong A_q$ and hence by Proposition~\ref{prop:quotcompr} $[0,1]\cong \pred(I) \cong \pred(A_q) \cong \downarrow q$. As a result $q$ is indeed proportional to an atomic effect. 
    
    For the other direction, use the universal property of filters (see \cite[197VII]{basthesis} for details) to write an effect $q$ as $q=g\circ\xi_{q}$ where $g$ is unital and $\xi_{q}:A\rightarrow A_q$. Our goal is to show that $g$ is an isomorphism so that $q$ is indeed pure. First of all, as $g:A_q\rightarrow I$ is unital, we have $g = \id_I\circ g = 1_I\circ g = 1_{A_q}$. Second, since $\downarrow q \cong [0,1]$ by assumption, Proposition~\ref{prop:quotcompr} gives us $\pred(A_q)\cong \downarrow q \cong [0,1] \cong \pred(I)$ so that $A_q$ is a scalar-like system. Hence by definition of an operational effect theory, there is an isomorphism $\Theta:I\rightarrow A_q$. Isomorphisms preserve the unit and hence $g\circ\Theta = 1_{A_q}\circ\Theta = 1_I = \id_I$, so that $g$ is the inverse of $\Theta$, and hence is an isomorphism. We conclude that $q$ is a composition of a filter and an isomorphism, and hence is pure.
    

	Using a similar sort of argument, when we have a pure state $\omega=\pi_s\circ \xi_t$ we must have $t=1$ so that $\xi_1$ is an isomorphism and then $\im{\omega}=\im{\pi_s}=s$ where $s$ must be proportional to an atomic effect because $[0,1]\cong \pred(I)\cong \pred(\{A\lvert s\}) \cong \downarrow s$.
\end{proof}

\begin{proposition} \label{prop:atomicstate}
	Let $q$ be an atomic effect. There exists a unique unital state $\omega_q$ such that $\im{\omega_q}=q$. This state is pure and given by $\omega_q:=q^\dagger$.
\end{proposition}
\begin{proof}
    First, for existence: since $q$ is atomic, it is pure by Proposition~\ref{prop:purestate}, hence $\omega_q:=q^\dagger$ exists and is a pure state. As also seen in that proof $q=\xi_q$, and hence $q^\dagger = \xi_q^\dagger = \pi_q$ by~\ref{pet:sharpadjoint}. Hence $\im{\omega_q}=\im{\pi_q} = q$. Furthermore, since compressions are unital (Proposition~\ref{prop:faithfulfilters}), $\omega_q$ is as well.

	For uniqueness we note that any unital state $\omega$ can be written as $\pi_{\im{\omega}}\circ \cl{\omega}$ where $\cl{\omega}$ is also a unital state and $\pi_{\im{\omega}}$ is a compression for $\im{\omega}$. Here $\cl{\omega}$ is a map to the object $\{A\lvert \im{\omega}\}$ which has $\pred(\{A\lvert \im{\omega}\}) \cong \downarrow \im{\omega}$ by Proposition~\ref{prop:quotcompr}. If $\im{\omega}=q$ is atomic then the effect space will be the real numbers so that $\{A\lvert \im{\omega}\}$ is a scalar-like system. Hence, using the same argumentation as in the proof of Proposition~\ref{prop:purestate}, $\cl{\omega}$ will be the unique unital state on this system. This means that any state with $\im{\omega}=q$ will be equal to $\pi_q$.
\end{proof}

\noindent We now have a correspondence between pure states and pure effects. A pure state $\omega$ has an atomic image $q=\im{\omega}$ that is a pure effect. It is the unique pure effect such that $q\circ\omega = 1$. In turn $\omega$ is the unique pure state for $q$ such that $q\circ\omega = 1$. We want this correspondence to satisfy the following property:

\begin{definition}
    We will say an operational PET has \textbf{symmetry of transition probabilities} \cite{alfsen2012geometry} when for any two atomic effects $p$ and $q$ on the same system we have $q\circ \omega_p= p\circ \omega_q$ where $\omega_p$ is the unique pure unital state with $\im{\omega_p}=p$.
\end{definition}

\noindent This is easy to show in the following case:
\begin{proposition}\label{prop:puredistinguish}
	Let $p$ and $q$ be atomic effects. We have $p\circ \omega_q = 0 \iff q\circ \omega_p = 0 \iff p\perp q$.
\end{proposition}
\begin{proof}
	$p\circ \omega_q=0 \iff p\leq \im{\omega_q}^\perp = q^\perp \iff p\perp q \iff q\perp p \iff q\leq \im{\omega_p}^\perp \iff q\circ \omega_p =0$.
\end{proof}

\noindent The general case is a bit harder to show, and we need the following lemma.

First, note that all scalars are pure maps as they are filters for themselves, and hence the dagger is defined for scalars.
\begin{lemma}\label{lem:scalarselfadjoint}
    Let $s: I\rightarrow I$ be a scalar, then $s^\dagger = s$.
\end{lemma}
\begin{proof}
    A scalar $s:I\rightarrow I$ of course corresponds to some number $s\in[0,1]$, and composition of scalars $s\circ t$ is then equal to their product $st$. We then have $(st)^\dagger = (s\circ t)^\dagger = t^\dagger \circ s^\dagger = s^\dagger t^\dagger$, so that the dagger preserves multiplication. 
    Note that for real numbers between $0$ and $1$ we have $s\leq t \iff \exists r\in[0,1]: s = rt$. As a consequence we get $s\leq t \iff s = rt \iff s^\dagger = r^\dagger t^\dagger \iff s^\dagger \leq t^\dagger$ so that the dagger is also an order-isomorphism for the unit-interval.
    Now suppose that $s\leq s^\dagger$, then by taking the dagger on both sides we get $s^\dagger \leq s$ so that $s=s^\dagger$. This of course also holds when we start with $s^\dagger \leq s$. Since the unit interval is totally ordered, one of these cases must be true and we are done.
\end{proof}

\begin{proposition}\label{prop:symoftrans}
    An operational PET has symmetry of transition probabilities.
\end{proposition}
\begin{proof}
    Let $p$ and $q$ be atomic effects. By Proposition~\ref{prop:purestate} they are then pure, and to be more specific they are equal to filters: $p=\xi_p:A\rightarrow I$, $q=\xi_q:A\rightarrow I$. Since $I$ only has one isomorphism on it, namely the identity, the filters for these effects with this type are unique. Similarly, from Proposition~\ref{prop:atomicstate} we see that their associated pure states $\omega_p$ and $\omega_q$ are equal to their unique compressions: $\omega_p = \pi_p$ and $\omega_q = \pi_p$. Using the dagger we then get $p^\dagger = \xi_p^\dagger = \pi_p = \omega_p$ and similarly $\omega_q^\dagger = q$.
    The expression $q\circ \omega_p$ is a scalar and hence by Lemma~\ref{lem:scalarselfadjoint} we have $q\circ \omega_p = (q\circ\omega_p)^\dagger = \omega_p^\dagger \circ q^\dagger = p\circ \omega_q$.
\end{proof}

\begin{definition}
    Let $v,w \in V$ be arbitrary vectors and write them as $v=\sum_i \lambda_i p_i$ and $w=\sum_j \mu_j q_j$, where the $p_i$ are orthogonal atoms and the same for the $q_j$. Such a decomposition can always be found by Corollary~\ref{cor:atomicspectrum}. Define the \textbf{inner product} of $v$ and $w$ to be $\inn{v,w} := \sum_{i,j} \lambda_i \mu_j (q_j\circ\omega_{p_i})$.
\end{definition}
\begin{proposition}
    The inner product defined above is indeed an inner product: well-defined, symmetric and $\inn{v,v}\geq 0$ with $\inn{v,v}=0$ iff $v=0$.
\end{proposition}
\begin{proof}
    First note that with $v$ and $w$ as defined above, 
    $$\inn{v,w} = \sum_{i,j} \lambda_i \mu_j (q_j\circ\omega_{p_i}) = \sum_i \lambda_i (\sum_j \mu_j q_j)\circ\omega_{p_i} = \sum_i \lambda_i (w\circ\omega_{p_i})$$ 
    so that the inner product is independent of the representation of $w$ in terms of atomic effects. 
    Second, due to symmetry of transition probabilities $q_j\circ\omega_{p_i} = p_i\circ\omega_{q_j}$ and hence $\inn{v,w} = \inn{w,v}$ so that it is also independent of the representation of $v$.

    Finally, we write $\inn{v,v} = \sum_{i,j} \lambda_i \lambda_j p_j\circ\omega_{p_i} = \sum_i \lambda_i^2 p_i\omega_{p_i} = \sum_i \lambda_i^2\geq 0$ due to Proposition~\ref{prop:puredistinguish}, and hence $\inn{\cdot,\cdot}$ indeed forms an inner product.
\end{proof}
\begin{proposition}
    The inner-product makes $V$ \textbf{self-dual}: i.e.\ we have $v\geq 0$ for $v\in V$ if and only if for all $w\geq 0$ we have $\inn{v,w}\geq 0$.
\end{proposition}
\begin{proof}
    If $v\geq 0$ then we can write $v=\sum_i \lambda_i p_i$ with the $p_i$ atomic and orthogonal and $\lambda_i \geq 0$ for all $i$. It then easily follows that $\inn{v,w}\geq 0$ if $w$ is also positive. For the other direction, suppose $\inn{v,w}\geq 0$ for all positive $w$, then in particular $\inn{v,p_i} = \lambda_i \geq 0$, so that $v$ is indeed positive.
\end{proof}
\begin{corollary} \label{cor:purestateconvex}
    Let $\omega\in $ St$_1(A)$ be a unital state, then $\omega = \sum_i \lambda_i \omega_{p_i}$ with $\lambda_i\geq 0$, $\sum_i\lambda_i = 1$ and the $\omega_{p_i}$ being pure states.
\end{corollary}
\begin{proof}
    The inner product defines a linear map $f:V\rightarrow V^*$ by $f(v)(w):= \inn{v,w}$. This map is an injection, so that due to finite-dimensionality it is a bijection. In particular, we can find for every $\omega \in V^*$ an element $v\in V$ such that $f(v)=\omega$ and hence $\omega(w) = \inn{v,w}$. Since $\omega(w)\geq 0$ for all $w\geq 0$ we must have $v\geq 0$. By expanding $v$ in terms of atomic effects we then get the desired result.
\end{proof}
\begin{corollary}\label{cor:stateconvexpure}
    A state is pure if and only if it is convex extremal.
\end{corollary}

\noindent The combination of a self-dual inner product and assert maps brings us very close to the setting of Ref.~\cite{alfsen2012geometry}. In particular by making the appropriate translations we can use their results to prove our main theorem:

\begin{theorem} \label{theor:OETEJA}
	Let $\mathbb{E}$ be an operational PET. Then there exists a functor ${F:\mathbb{E}\rightarrow \EJA^\opp_{\text{psu}}}$ such that for any system $A$ the effect space $\pred(A)$ is isomorphic to the unit interval of its corresponding EJA: $\pred(A)\cong [0,1]_{F(A)}$. Furthermore, this functor is faithful if and only if $\mathbb{E}$ is locally tomographic.
\end{theorem}
\begin{proof}
    By Theorem~\ref{theor:opefftheor} we have a functor $F:\mathbb{E}\rightarrow \OUS$, so that it suffices to show that the order unit spaces that all systems are mapped to are actually EJAs. We show this by establishing all the conditions of Theorem 9.33 of \cite{alfsen2012geometry}. Most of the work for this has already been done, but some technical conditions still need to checked. The details can be found in Appendix \ref{sec:proofofeja}. As in Theorem~\ref{theor:opefftheor}, faithfulness of the functor corresponds to local tomography in $\mathbb{E}$.
\end{proof}


\section{Monoidal Operational PETs}\label{sec:quantum}
We have now shown that systems in an operational PET correspond to Euclidean Jordan algebras. Conversely, the results of Ref.~\cite{westerbaan2018puremaps} show that the opposite category of EJAs with positive sub-unital maps is an operational PET. Hence, we have characterized operational PETs.

In this section we will see what additional restrictions are imposed on the systems by adding a monoidal structure to the theory that respects the structure of pure maps.

The canonical example of an monoidal operational PET is \textbf{CStar}$^\opp_{\text{cpsu}}$, the opposite category of complex finite-dimensional C$^*$-algebras with completely positive sub-unital maps. Another example is \textbf{RStar}$^\opp_{\text{cpsu}}$ of \emph{real} finite-dimensional C$^*$-algebras with completely positive sub-unital maps. A real finite-dimensional C$^*$-algebra is a direct sum of the real matrix algebras $M_n(\R)$. It turns out that these two examples are the only possibilities.

\begin{theorem}\label{theor:compositealgebras}
    Let $\mathbb{E}$ be an monoidal operational PET. Then there is a functor $F:\mathbb{E}\rightarrow \mathbb{D}$ where $\mathbb{D} = \textbf{CStar}^\opp_{\text{cpsu}}$ or $\mathbb{D} = \textbf{RStar}^\opp_{\text{cpsu}}$. This functor preserves the effect space of objects: $\pred(A)\cong [0,1]_{F(A)}$. 
\end{theorem}
\begin{proof}
    By Theorem~\ref{theor:OETEJA} we have a functor $F:\mathbb{E}\rightarrow \EJA^\opp_{\text{psu}}$ so that it suffices to show that the image of this functor lands in the real or complex C$^*$-algebras.

    We give a sketch of the proof here, for the details we refer to Appendix~\ref{sec:compositesappendix}. Note that this proof closely follows the structure of proofs showing how to go from arbitrary EJAs to C*-algebras in Refs.~\cite{wetering2018sequential,selby2018reconstructing}.

    First we show that composites of atomic effects are again atomic and that the composite preserves orthogonality of atoms. As a result we establish that the composite acts as you would expect with regard to the rank and dimension of the space. We use the argument from Ref.~\cite{wetering2018sequential} to show that composites of simple EJAs must again be simple. A simple dimension-counting argument establishes that the only simple factors allowed are the real and complex matrix algebras. The final step is to show that it is impossible to form a composite of a real and a complex matrix algebra that behaves correctly with regard to pure maps.
 \end{proof}

 \begin{remark}
    This theorem does not mention the faithfulness of this functor. If $\mathbb{E}$ satisfies local tomography, then in the same way as in Theorem~\ref{theor:OETEJA} we see that it is faithful, but we expect that faithfulness already holds if $\mathbb{E}$ merely satisfies tomography. To determine this however we would need to show that the functor $F$ is strongly monoidal which does not seem trivial. We also expect that if $\mathbb{E}$ satisfies local tomography that the effect spaces of $\mathbb{E}$ must be isomorphic to those of complex instead of real C$^*$-algebras, but again, to prove this it seems like we need to know more about $\mathbb{E}$. We leave this as an open question for future work.
 \end{remark}

\begin{remark}
It should be noted that $\textbf{vNA}^\opp_{\text{cpsu}}$, the opposite category of von Neumann algebras with completely positive sub-unital maps, is also a monoidal PET~\cite{bramthesis,basthesis}, but since it contains infinite-dimensional system it is not an operational PET. Finding suitable conditions under which we retrieve $\textbf{vNA}^\opp_{\text{cpsu}}$ (or the bigger category of JBW-algebras and positive maps) is still an open question.
\end{remark}

\acknowledgments{
The author would like to thank Bas and Bram Westerbaan for all the useful and insightful conversations regarding effect algebras and order unit spaces and also for spotting some errors in the original proofs. Thanks also go out to Aleks Kissinger for helpful suggestions regarding the presentation of the work. Finally, the author thanks the anonymous reviewers for their helpful comments, in particular one reviewer who pointed out some flaws in the original use of (local) tomography. This work was supported by the ERC under the European Union’s Seventh Framework Programme (FP7/2007-2013) / ERC grant n$^\text{o}$ 320571 and by AFOSR grant FA2386-18-1-4028.
}

\appendix

\section{Operational Effect Theories}\label{sec:opefftheory}

It is a matter of routine to check that the opposite category $\OUS$ forms an effect theory where the effects correspond to elements in $[0,1]_V$. Now since the states living in the continuous dual space $V^*$ order-separate $V$ \cite[Corollary 1.27]{alfsen2012state} we see that $\OUS$ in fact forms an operational effect theory.

\begin{proposition}
    Let $\mathbf{E}$ be an effect theory with real scalars and where the unital states order-separate the effects. Then there are no infinitesimal effects, i.e.~for $p,q\in\pred(A)$, if $\frac{1}{2}p\leq \frac{1}{2}(q+1/n)$ for all $n\in\N_{>0}$ then $p\leq q$.
\end{proposition}
\begin{proof}
    Let $p$ be an effect such that $\frac{1}{2}p\leq \frac{1}{2}(q+1/n)$ for all $n\in \N_{>0}$. Then for all unital states $\omega$ we will have $(\frac{1}{2}p)\circ \omega\leq \frac{1}{2}(q+1/n)\circ \omega = (\frac{1}{2}q)\circ\omega + 1/(2n)(1\circ \omega)$. 
    As this must hold for all $n$ we conclude that $(\frac{1}{2}p)\circ \omega\leq (\frac{1}{2}q)\circ \omega$. 
    By assumption the unital states order-separate the effects so that we conclude that $\frac{1}{2}p\leq \frac{1}{2}q$ and hence $p\leq q$.
\end{proof}

\begin{definition}
    We call an effect algebra $E$ an \textbf{effect module} (better known as a \textbf{convex effect algebra}) when there is an action of the real unit interval $[0,1]$ on $E$ such that $1\cdot e = e$, $0\cdot e = 0$, $(x+y)\cdot e = x\cdot e + y\cdot e$ and $(xy)\cdot e = x\cdot (y\cdot e)$. Such an action is necessarily unique~\cite{gudder1999convex}. We let \textbf{Emod} denote the category of effect modules with maps that preserve addition and scalar multiplication, and \textbf{Emod}$_a$ the full subcategory of \emph{Archimedean} effect modules, i.e.~those that do not have infinitesmal effects as defined in the previous proposition.
\end{definition}

In an operational effect theory the sets of effects form effect modules, because when we have an effect $q:A\rightarrow I$ and a scalar $s: I\rightarrow I$, we get a new effect $s\circ q$ by composition. It can easily be checked that this action satisfies the properties necessary to be an effect module.

\begin{proposition}
\cite{gudder1999convex,jacobs2016expectation} There is a one-to-one correspondence between Archimedean effect modules and order unit spaces: if $E$ is an Archimedean effect module, then we can find an order unit space $V$ such that $E\cong [0,1]_V$.
\end{proposition}

\begin{proof}[\textbf{Proof of Theorem~\ref{theor:opefftheor}}]
    There is the obvious effects functor for an effect theory {Eff:$\mathbb{E} \rightarrow $ \textbf{Emod}$^\opp$} that maps each object to its set of effects and transformations $f:A\rightarrow B$ to $f^*: \pred(B)\rightarrow \pred(A)$ given by $f^*(q) = q\circ f$. In an operational effect theory the sets of effects form Archimedean effect modules so that the image of this functor lands in \textbf{Emod}$_a$. We know by Ref.~\cite{jacobs2016expectation} that there is an equivalence of categories $F:$ \textbf{Emod}$_a\rightarrow $\textbf{OUS}, so that $F^\opp~\circ$ Eff $:\mathbb{E} \rightarrow \OUS$. It is then straightforward to see that the faithfulness of this functor exactly corresponds to local tomography of $\mathbb{E}$.
\end{proof}

\section{Proof of Theorem \ref{theor:OETEJA}} \label{sec:proofofeja}

We will use a theorem by Alfsen and Shultz (\cite[Theorem 9.33]{alfsen2012geometry}) to prove Theorem~\ref{theor:OETEJA}. Their proof relies on a type of operators they also call \emph{compressions}. Their compressions will turn out to be our assert maps. To distinguish our compressions from theirs, we will call their compressions \emph{AS-compressions}.

\begin{definition}
    Let $P:A\rightarrow A$ be a map in an effect theory. We call it an \textbf{AS-compression} when $P$ is idempotent and it is \textbf{bicomplemented}: there exists an idempotent $Q:A\rightarrow A$ such that for all effects $q$ and states $\omega$ the following implications hold:
    \begin{itemize}
        \item $q\circ P = q \iff q\circ Q = 0$.
        \item $q\circ Q = q \iff q\circ P =0$.
        \item $P\circ \omega = \omega \iff Q\circ \omega = 0$.
        \item $Q\circ \omega = \omega \iff P\circ \omega =0$.
    \end{itemize}
\end{definition}

\begin{proposition}\label{prop:gencompr}
    In a PET where the effects separate the states (i.e.\ where $\omega=\omega^\prime$ when $p\circ \omega = p\circ \omega^\prime$ for all effects $p$) the assert map $\asrt_p$ of a sharp effect $p$ is an AS-compression with bicomplement $\asrt_{p^\perp}$.
\end{proposition}
\begin{proof}
    In Proposition~\ref{prop:assertmaps} it was already shown that assert maps of sharp effects are idempotent so it remains to show that they are bicomplemented. The bicomplement of $\asrt_p$ will turn out to be $\asrt_{p^\perp}$.

    We have $q=q\circ \asrt_p \iff q\leq p=(p^\perp)^\perp \iff q\circ \asrt_{p^\perp}=0$ by an application of Propositions \ref{prop:assertmaps} and \ref{prop:orthog}. Obviously the same holds with $p$ and $p^\perp$ interchanged.

    Let $\omega$ be a state. Suppose $\asrt_p\circ \omega = \omega$ then $q\circ \asrt_{p^\perp}\circ \omega = q\circ\asrt_{p^\perp}\circ \asrt_p\circ\omega \leq p^\perp\circ \asrt_p\circ \omega = 0\circ\omega = 0$ for all effects $q$. Because the effects separate the states we can conclude that $\asrt_{p^\perp}\circ \omega = 0$. Now for the other direction suppose $\asrt_{p^\perp}\circ \omega = 0$, then $0=1\circ \asrt_{p^\perp}\circ \omega = p^\perp\circ \omega$ so that $1\circ \omega = (p+p^\perp)\circ \omega = p\circ \omega$. Then by definition of images: $\im{\omega}\leq p$ so that by Proposition \ref{prop:assertmaps} $\asrt_p\circ \omega = \omega$. The other direction we get by interchanging $p$ and $p^\perp$. So $\asrt_p$ and $\asrt_{p^\perp}$ are indeed bicomplemented.
\end{proof}

The results of Ref.~\cite{alfsen2012geometry} require a few concepts from convex geometry, in particular the notion of a \emph{face}. Even though the definition we give below might seem foreign, it in fact corresponds with the intuitive idea of an extreme face of a convex set.

\begin{definition}
    Let $W$ be a real vector space and let $K\subseteq W$ be a convex subset. A \textbf{face} $F$ of $K$ is a convex subset such that whenever $\lambda x + (1-\lambda) y\in F$ with $0<\lambda<1$ then $x, y \in F$. Any extreme point of $K$ forms a face on its own. A face is called \textbf{norm exposed} when there exists a bounded affine positive functional $f:K\rightarrow \R^+$ such that $f(a)=0\iff a\in F$. A face is called \textbf{projective} when there exists an AS-compression $P:W\rightarrow W$ such that for all $x\in K$ we have $P(x)=x \iff x\in F$.
\end{definition}

\begin{lemma}\label{lem:normexposed}
	Let $A$ be a system in an operational PET. Any norm-exposed face of St$(A)$ is projective.
\end{lemma}
\begin{proof}
	Let $V$ be the vector space associated to $A$. Let $F\subseteq $ St$(A)$ be a norm-exposed face. That means that there is a positive affine functional $f:$ St$(A)\rightarrow \R_{\geq 0}$ with $f(\omega)=0\iff \omega\in F$. As St$(A)$ is the state space of an order unit space, it forms a \emph{base} of the \emph{base norm space} $V^*$~\cite[Theorem 1.19]{alfsen2012state}. As a result, its span is the entire positive cone of $V^*$~\cite[Definition 1.10]{alfsen2012state}, and $f$ extends uniquely to a positive linear map $f:V^*\rightarrow \R$~\cite[Proposition 1.11]{alfsen2012state}. In finite dimension we of course have $V\cong V^*$ so that there must be a $q\geq 0$ in $V$ such that $\forall \omega\in$ St$(A): f(\omega) = \omega(q)$. We can rescale $q$ without changing the zero set, so we can take $q$ to be an effect. By Proposition~\ref{prop:floorceiling} $\omega(q)=0\iff \omega(\ceil{q})=0$, so $q$ can be replaced by a sharp effect without changing the zero set. Now $\omega \in F \iff \omega(q)=0 \iff $ $\im{\omega}\leq q^\perp \iff \asrt_{q^\perp}\circ \omega = \omega$. Since assert maps of sharp effects are AS-compressions we see that $F$ is indeed projective.
\end{proof}

\begin{corollary} \label{cor:comp-assert}
	Let $A$ be a system in an operational PET. Any AS-compression is an assert map.
\end{corollary}
\begin{proof}
	If we have a AS-compression $P$ with complement $Q$ then we can construct $f:$ St$(A)\rightarrow \R^+$ by $f(\omega) = \norm{Q(\omega)}$ which is affine (this is a standard result for the norm on base norm spaces in order separation with an order unit space \cite{alfsen2012geometry}). Now obviously $f(v)=0 \iff Q(v)=0 \iff P(v)=v$, so we see that the projective face generated by $P$ is also norm exposed. But then by the previous proposition it is also the projective face of some assert map which necessarily must have the same projective unit. Because the projective unit determines the compression uniquely we see that $P$ must be equal to this assert map.
\end{proof}

\begin{lemma}\label{lem:compressionpure}
	In an operational PET, the AS-compression of a convex extremal state is proportional to a convex extremal state.
\end{lemma}
\begin{proof}
    By Corollary~\ref{cor:stateconvexpure} convex extremal states are precisely the pure states.
	The only AS-compressions are the assert maps of sharp effects, and assert maps are pure maps. By \ref{pet:dagger} the composition of pure maps is again pure, so that an AS-compression sends a convex extremal state to a pure state, which again by Corollary~\ref{cor:stateconvexpure} must be convex extremal.
\end{proof}

Theorem 9.33 of \cite{alfsen2012geometry} states that a state-space is isomorphic to that of a Jordan algebra when the following conditions are met:
\begin{itemize}
    \item Every norm exposed face is projective (Lemma \ref{lem:normexposed}).
    \item The convex extremal points span the space (Corollary \ref{cor:purestateconvex}).
    \item It satisfies symmetry of transition probabilities (Proposition \ref{prop:symoftrans}).
    \item AS-compressions preserve convex extremal states (Lemma \ref{lem:compressionpure}).
\end{itemize}

Hence, we can indeed conclude that our systems are Euclidean Jordan algebras.

\section{Proof of Theorem~\ref{theor:compositealgebras}}\label{sec:compositesappendix}

As it has already been established that the effect spaces of systems in a operational PET correspond to Euclidean Jordan algebras, we will let $V$ and $W$ denote EJAs that should be understood to come from an operational PET. The monoidal structure of the PET obviously lifts to a bilinear map $V\times W\rightarrow V\otimes W$ where $V\otimes W$ is some other EJA corresponding to a system of the PET (not necessarily related to the regular vector space tensor product). In particular, if we have $v\geq 0$ in $V$ and $w\geq 0$ in $W$ then $v\otimes w\geq 0$ in $V\otimes W$.

\begin{proposition} \label{prop:compositeproperties}
    In a monoidal operational PET the following are true.
    \begin{enumerate}
        \item A composite of pure maps is again pure.
        \item A composite of normalized states is again a normalized state.
        \item A composite of atomic effects is again an atomic effect.
        \item For atomic effects $p$ and $q$ we have $\omega_p\otimes \omega_q = \omega_{p\otimes q}$.
        \item If $q_1\perp p_1$ and $q_2\perp p_2$ are atomic orthogonal effects, then $q_1\otimes q_2 \perp p_1\otimes p_2$.
    \end{enumerate}
\end{proposition}
\begin{proof}~
    \begin{enumerate}
        \item By definition of a monoidal PET.
        \item Given two normalized states $\omega_1$ and $\omega_2$ we see that $1\circ(\omega_1\otimes \omega_2) = (1\otimes 1)\circ(\omega_1\otimes \omega_2) = (1\circ\omega_1) \otimes (1\circ \omega_2) = 1\otimes 1 = 1$.
        \item Let $p$ and $q$ be atomic effects. By Proposition \ref{prop:purestate} this is equivalent to them being sharp and pure. By point 1 we know that $p\otimes q$ is also pure and hence $p\otimes q$ must be proportional to an atom: $p\otimes q = \lambda r$. We calculate $1 = (p\otimes q)\circ (\omega_p\otimes \omega_q) = (\lambda r)\circ (\omega_p\otimes \omega_q) \leq (\lambda 1)\circ (\omega_p\otimes \omega_q) = \lambda$, and hence $\lambda=1$ and we are done.
        \item We know that $p\otimes q$ is atomic, and we know that $\omega_p\otimes \omega_q$ is a unital pure state (since it is a composite of pure unital states). We of course have $(p\otimes q)\circ (\omega_p\otimes \omega_q) = 1$, but by Proposition \ref{prop:atomicstate}, the state $\omega_{p\otimes q}$ is the unique state with this property and hence $\omega_{p\otimes q} = \omega_p\otimes \omega_q$.
        \item By Proposition \ref{prop:puredistinguish}, atomic effects $p$ and $q$ are orthogonal if and only if $q\circ \omega_p = 0$. So supposing that $q_1\perp p_1$ and $q_2\perp p_2$ we calculate $(q_1\otimes q_2)\circ\omega_{p_1\otimes p_2} = (q_1\otimes q_2)\circ (\omega_{p_1}\otimes \omega_{p_2}) = (q_1\circ \omega_{p_1})\otimes (q_2\circ \omega_{p_2}) = 0\otimes 0 = 0$. \qedhere
    \end{enumerate}
\end{proof}

\begin{definition}
    The \textbf{rank} of an EJA $V$, denoted by $\rnk~V$, is equal to the maximal size of any set of orthogonal atomic effects in $V$.
\end{definition}
\noindent The rank of the $n\times n$ matrix algebra $M_n^{sa}(F)$ is equal to $n$. The rank of a spin-factor is always equal to 2. Note that the size of a set of of orthogonal atomic effects $\{p_i\}$ is equal to the rank of the space if and only if $\sum_i p_i = 1$.

\begin{proposition}\label{prop:rankcomposite}
    Rank is preserved by compositing: $\rnk~V\otimes W = \rnk~V \rnk~W$. We also have $\dim~V\otimes W \geq \dim~V \dim~W$.
\end{proposition}
\begin{proof}
    Let $\{p_i\}$ be a maximal set of orthogonal atomic effects in $V$, and let $\{q_j\}$ be a maximal set of orthogonal atomic effects in $W$. By maximality we must have $\sum_i p_i = 1_V$ and $\sum_j q_j = 1_W$. By Proposition \ref{prop:compositeproperties} the set $\{p_i\otimes q_j\}$ also consists of orthogonal atomic effects. Furthermore $\sum_{i,j} p_i\otimes q_j = (\sum_i p_i)\otimes (\sum_j q_j) = 1_V\otimes 1_W = 1_{V\otimes W}$ so that this set must also be maximal.

    For the second part let $\{p_i\}$ be a basis of atomic effects of $V$ and similarly let $\{q_j\}$ be a basis of atomic effects in $W$. Suppose that $\dim~V\otimes W < \dim~V \dim~W$, then $\{p_i \otimes q_j\}$ must be linearly dependent in $V\otimes W$. We will show that this leads to a contradiction.
    If this set is linear dependent, than without loss of generality we can write $p_1\otimes q_1 = \sum_{i,j} \lambda_{ij} p_i\otimes q_j$, for some real numbers $\lambda_{ij}$ where the sum goes over all $i,j$ except $i=j=1$. 
    Let $\omega$ be an arbitrary unital state on $W$ and apply the map $\id\otimes \omega$ to both sides to get $p_1\otimes \omega(q_1) = \omega(q_1) p_1 = \sum_{i,j} \lambda_{ij} p_i \omega(q_j) = \sum_i p_i (\sum_j \lambda_{ij}\omega(q_j)) = p_1(\sum_j \lambda_{1j} \omega(q_j)) + \sum_{i>1} p_i (\sum_j \lambda_{ij}\omega(q_j)$.
    Rewrite this to $p_1(\omega(q_1)-\sum_j \lambda_{1j} \omega(q_j)) = \sum_{i>1} p_i \sum_j \lambda_{ij}\omega(q_j)$. Since by assumption the $p_i$ are linearly independent this can only hold when $\sum_j \lambda_{ij} \omega(q_j) = \omega(\sum_j \lambda_{ij} q_j) =  0$ for all states $\omega$ and $i>1$ and hence we must have $\sum_j \lambda_{ij} q_j = 0$. Since the $q_j$ are also linearly independent this shows that $\lambda_{ij} = 0$ when $i>1$. By interchanging the role of $p_1$ and $q_1$ we also get $\lambda_{ij} = 0$ when $j>1$, so that the only nonzero value could be $\lambda_{11}$, which finishes the contradiction.
\end{proof}

\begin{proposition}\label{prop:simplecomposite}
    If $V$ and $W$ are simple than their composite $V\otimes W$ is also simple.
\end{proposition}
\begin{proof}
    We know that the composite $V\otimes W$ has rank $\rnk~V \rnk~W$, so if we can show that $V\otimes W$ must contain a simple factor of this rank than we are done.

    Let $\{p_i\}$ be a maximal orthogonal set of atoms in $V$ and let $\{q_j\}$ be a maximal orthogonal set of atoms in $W$. Let $p$ be an atom in $V$ such that $\omega_p(p_i) \neq 0$ for all $i$, and similarly let $q$ be in $W$ such that $\omega_w(w_j) \neq 0$ (that such an atom can always be found can be straightforwardly verified by doing a case distinction on the possible simple factors). Then $(\omega_{p\otimes q})(p_i\otimes q_j) = \omega_p(p_i)\omega_q(q_j) \neq 0$. As a result we conclude that $p\otimes q$ must belong to the same simple factor as all the $p_i\otimes q_j$, and hence this simple factor must have rank at least $\rnk~V \rnk~W$.
\end{proof}

\begin{proposition}\label{prop:simpleEJAcomplexreal}
    Let $V$ be simple, then $V=M_n^{sa}(F)$ with $F=\R$ or $F=\C$.
\end{proposition}
\begin{proof}
    This can be shown by a simple case distinction and dimension counting argument. We will work out one specific case, the other ones follow similarly. Suppose $V=M_n^{sa}(\mathbb{H})$. By Propositions \ref{prop:rankcomposite} and \ref{prop:simplecomposite} we then know that $V\otimes V$ must be a simple EJA with rank $n^2$ and $\dim(V\otimes V)\geq \dim(V)^2$. The simple EJA of rank $n^2$ with the highest dimension is $M_{n^2}^{sa}(\mathbb{H})$. When $n>1$ however, the dimension of this space is still lower than $\dim(M_n^{sa}(\mathbb{H}))^2$, and hence such an EJA does not exist.

    Note that the spin-factors $S_2$ and $S_3$ do allow the right sort of composites, but that these are in fact isomorphic to respectively real and complex matrix algebras.
\end{proof}

\begin{proposition}\label{prop:complexrealexclusion}
    Let $V$ and $W$ be simple EJAs, then both $V$ and $W$ are real matrix algebras, or both of them are complex matrix algebras.
\end{proposition}
\begin{proof}
    By the previous proposition we know that they both must be real or complex, so that the only thing we need to show is that it cannot be that $V$ is complex while $W$ is real.

    Let $V=M_n^{sa}(\C)$ and $W=M_m^{sa}(\R)$. By dimension counting and the previous proposition we know that $V\otimes W = M_{nm}^{sa}(\C)$. Let $\omega$ be a pure state on $V$. The identity map on $W$ is of course pure, so that $\omega\otimes \id: M_{nm}^{sa}(\C) \rightarrow M_m^{sa}(\R)$ is also a pure map. The study of pure maps on EJAs in~\cite{westerbaan2018puremaps} shows however that all quotient and comprehension spaces of EJAs are isomorphic to certain subalgebras of the space. In particular, if $f: M_k^{sa}(\C)\rightarrow W$ is a pure map, then $W\cong M_l^{sa}(\C)$. The only time when $M_m^{sa}(\R) \cong M_l^{sa}(\C)$ is when $m=l=1$, and hence the proposition is proved.
\end{proof}

Finally, we can prove Theorem~\ref{theor:compositealgebras}.
Let $A$ be a system in a monoidal operational PET. By Theorem~\ref{theor:OETEJA} we know that $\pred(A)$ is isomorphic to the unit interval of a Euclidean Jordan algebra. Let $c$ be a minimal central element of this EJA, that hence corresponds to a simple factor of the algebra. The filter space associated to $c$ is then a simple algebra, and hence by Proposition~\ref{prop:simpleEJAcomplexreal} it must be a real or complex matrix algebra. let $d$ be another minimal central element. Then its associated simple factor is also a real or complex matrix algebra and furthermore this algebra is associated to some system in the PET. By Proposition~\ref{prop:complexrealexclusion} we can then conclude that either both the simple factors associated to $c$ and $d$ are real, or they are complex. Since this holds for all the simple factors we conclude that indeed Eff$(A)$ is isomorphic to the unit interval of a real or complex C*-algebra. That all the systems in the OET must be all real or all complex follows similarly.
Hence the functor of Theorem~\ref{theor:OETEJA} restricts to the category of real or complex C$^*$-algebras.


\bibliographystyle{plainnat}
\bibliography{bibliography}

\end{document}